\documentclass[12pt,apj]{emulateapj}

\usepackage{natbib, graphics}

\usepackage{longtable}
\usepackage{amsfonts}
\usepackage{rotating}
\usepackage{float}
\usepackage{threeparttable}

\RequirePackage{amssymb}
\RequirePackage{latexsym}
\RequirePackage{graphicx}

\renewcommand{\deg}{\mbox{$^{\circ}$}}

\newcommand{\omcen}{\mbox{$\omega$ Cen}} 
\newcommand{\msun}{\mbox{$M_{\odot}$}}
\newcommand{\strom}{\mbox{Str\"omgren~}}
\newcommand{\omc}{\mbox{$\omega$ Cen~}}

\begin{document}

\title{The not so simple globular cluster \omcen. I. Spatial distribution of the multiple stellar populations.\altaffilmark{1}}

\shorttitle{The not so simple globular cluster \omcen} 

\shortauthors{Calamida et al.}

\author{A. Calamida$^{2,3}$,
G. Strampelli$^4$,
A. Rest$^4$, 
G. Bono$^{5,3}$,
I. Ferraro$^3$,
A. Saha$^2$,
G. Iannicola$^3$, 
D. Scolnic$^6$,
D. James$^{7,8}$,
C. Smith$^7$,
A. Zenteno$^7$}

\altaffiltext{1} {Based on observations made with the Dark Energy Camera (DECam) on the 
4m Blanco telescope (NOAO) under programs 2014A-0327, 2015A-0151, 2016A-0189, PIs: A. Calamida, A. Rest,
and on observations made with the NASA/ESA Hubble Space Telescope,
obtained by the Space Telescope Science Institute. STScI is operated by the
Association of Universities for Research in Astronomy, Inc., under NASA
contract NAS 5-26555.}

\altaffiltext{2}
{National Optical Astronomy Observatory - AURA,
950 N Cherry Ave,
Tucson, AZ, 85719, USA;
calamida@noao.edu}

\altaffiltext{3}
{INAF - Osservatorio Astronomico di Roma - Via Frascati 33, 00040, Monteporzio Catone, Rome, Italy}

\altaffiltext{4}
{Space Telescope Science Institute - AURA, 
3700 San Martin Dr.,
Baltimore, MD 21218, USA}

\altaffiltext{5}
{Dipartimento di Fisica, Universit\'a di Roma Tor Vergata, Via della Ricerca Scientifica 1, 000133, Roma, Italy}

\altaffiltext{6}
{The University of Chicago,The Kavli Institute for Cosmological Physics, 
William Eckhardt Research Center - Suite 499,
5640 South Ellis Avenue,
Chicago, IL 60637, USA}

\altaffiltext{7}
{Cerro Tololo Inter-American Observatory, Casilla 603, La Serena, Chile}

\altaffiltext{8}
{Astronomy Department, University of Washington, Box 351580, Seattle, WA 98195, USA}

\begin{abstract}
We present a multi-band photometric catalog of $\approx$ 1.7 million cluster members 
for a field of view of $\approx$ 2$\times$2$\deg$ across \omcen.
Photometry is based on images collected with the Dark Energy Camera 
on the 4m Blanco telescope and the Advanced Camera for Surveys on the Hubble 
Space Telescope. The unprecedented photometric accuracy and field coverage 
allowed us for the first time to investigate the spatial distribution 
of \omc multiple populations from the core to the tidal radius, confirming 
its very complex structure.
We found that the frequency of blue main-sequence stars is 
increasing compared to red main-sequence stars starting from a 
distance of $\approx$ 25' from the cluster center. Blue main-sequence stars also 
show a clumpy spatial distribution, with an excess in the North-East quadrant of 
the cluster pointing towards the direction of the Galactic center.
Stars belonging to the reddest and faintest red-giant branch also show a more 
extended spatial distribution in the outskirts of \omcen, a region 
never explored before. 
Both these stellar sub-populations, according to spectroscopic measurements, 
are more metal-rich compared to the cluster main stellar population.
These findings, once confirmed, make \omc the only stellar system currently known 
where metal-rich stars have a more extended spatial distribution compared to metal-poor stars.
Kinematic and chemical abundance measurements are now needed for stars in the external regions of \omc
to better characterize the properties of these sub-populations.
\end{abstract}


\maketitle

\section{Introduction}\label{intro}
The peculiar Galactic Globular Cluster (GGC) \omc (NGC\,5139) has been
subject to substantial observational efforts covering 
the whole wavelength spectrum from the ultraviolet to the near-infrared. This gigantic star cluster, the most 
massive known in our Galaxy, $M = 2.5 \times 10^6$ \msun \citep{vandeven2006} 
has (at least) three separate stellar populations with a large undisputed spread in
metallicity \citep{norris_dacosta1995, norris1996, suntzeff1996, 
kayser2006, villanova2007, calamida2009, johnson2010}.
Table 1 summarizes the basic parameters of \omcen.

It has been suggested that \omc stellar populations not only show
different chemical abundances but also have different kinematical properties.
In particular, \citet{norris1997} matching the spectroscopic abundances of $\approx$ 500 
red-giant (RG) stars with the radial velocities by \citet{mayor1997}, observed that the 
metal-rich (MR) stars of \omc do not share the rotational velocity (V $\approx$ 8 Km s$^{-1}$) 
of the metal-poor (MP) component. The most MR stars also seem to have a smaller 
velocity dispersion compared to the MP stars. However, these results were questioned by \citet{pancino2007} and 
\citet{sollima2009}, who found that the most MR stellar component of \omc does not present 
any significant radial velocity offset with respect to the bulk of stars. 
Moreover, the velocity dispersion profile appears to decrease monotonically 
from $\sigma_v \approx 17.2$ km/s down to a minimum value of $\sigma_v \approx 5.2$ km/s, 
in the region 1.5 $\le r \le$ 28'. For distances larger than 30' an hint of a raise
in the velocity dispersion is present, but this result is not statistically 
significant \citep{sollima2009}. 

A study of proper motions of the different cluster stellar populations was performed 
by \citet{ferraro2002}, who combined the proper motion data of \citet{vanLeeuwen2000} 
with the photometry of \citet{pancino2000}. This analysis showed that the most MR stars in the cluster
have a different motion compared to the metal-intermediate (MI) and MP stars. They suggested
that the most MR stars in \omc formed in an independent stellar system later 
accreted by the cluster.

\citet{pancino2000, pancino2003}, by analyzing the spatial distribution of 
a sample of cluster RG stars, concluded that the three main stellar populations 
of \omc (MP, MI, and MR) have different 
distributions: MP stars are distributed along the direction of the cluster major axis
(E--W), while the MI and MR along the N--S axis. This result was confirmed by \citet{hilker2000},
based on their \strom photometric metallicities for a sample of \omc RGs. In particular, 
they found that the more MR stars seem to be more concentrated within a radius 
of 10\arcmin~ from the cluster center. 
\citet{sollima2005a} also found that MR stars
are more centrally concentrated based on photometry of RG 
stars for a field of view (FoV) of $\approx$ 0.2$\times$0.2$\deg$ across the cluster.

Another peculiar property of \omc is the splitting of the main-sequence (MS).  
Hubble Space Telescope (HST) photometry revealed
for the first time that \omc MS bifurcates into two main components, the so called blue-MS (bMS) 
and red-MS \citep{anderson2002,bedin2004}. Spectroscopic follow-up by \citet{piotto2005} showed 
that bMs stars are more metal-rich than rMS stars, while their color being bluer. 
These authors then suggested that bMS stars constitute a helium-enhanced sub-population in the cluster.
The split of \omc MS was also found by \citet{sollima2007a} based on Very Large Telescope (VLT) photometry.
They showed that the two sequences are still well-separated at $\approx$ 26\arcmin~ from the cluster center and that the bMS is more
centrally concentrated compared to the rMS. 

The formation history and composition of \omc form a complex puzzle 
that is being slowly pieced together by investigations based on the latest 
generation of telescopes. 
However, all previous photometric studies were based on catalogs covering a field
of view of no more than $\approx$ 40$\times$40\arcmin~ across \omcen. 
We now push forward the ongoing investigations by combining precise multi-band DECam 
photometry, covering $\approx$ 2$\times$2$\deg$ across the cluster, and HST data for the cluster 
center to characterize the properties of \omc multiple stellar populations from 
the core to the tidal radius.

The structure of the current paper is as follows. In \S 2 we discuss 
the observations and data reduction of DECam and ACS data of \omcen. In \S 3 
we illustrate the calibration strategy used for the ground-based data set 
and \S 4 deals with the selection criteria 
adopted to identify candidate field and cluster stars. In \S 5 we 
present the ground-based color-magnitude diagrams while in \S 6 we 
illustrate how we performed the astronometric calibration of our data set. 

In \S 7 and 8 we discuss the spatial distribution of the different 
\omc stellar populations and we summarize and discuss the results in \S 9.
 
\section{Observations and data reduction}\label{obs}
Photometric data discussed in this investigation belong to two different 
sets from both space and ground based telescopes. 
A set of 57 {\it ugri}  images centered on \omc 
was collected over 3 nights, 2014 February 24, 2015 June 22 and 2016 March 4, 
with the Dark Energy Camera (DECam) on the 4m Blanco Telescope (CTIO, NOAO). 
DECam is a wide-field imager with 62 CCDs and covers a 3 square degree sky FoV
with a pixel scale of 0.263\arcsec.
Exposure times for our observations ranged from 120 to 600s for the $u$-band and from 7 to 250s for the other filters.
Weather conditions were very good for all nights with image seeing ranging 
from 0.8\arcsec~ to 1.6\arcsec~ for the $u$-band and from 0.7\arcsec~ to 1.2\arcsec~ 
for the other filters. Standard stars from the Sloan Digital Sky Survey (SDSS) 
Stripe 82 were observed in all filters and at different air masses for the 
night of February 2014. The accuracy of the derived zero points (ZPs)
ranges between 2\% for the $r$ and $i$ filters to $4-5$\% for the $g$ and
$u$ filters. For more details on the photometric calibration see Table~3 and Section \S 3.
Table~2 lists the log of the DECam observations while 
Fig.~1 shows the footprint of DECam photometric catalog (black dots) for \omcen.
Note that stars are missing at the top and bottom of the footprint since 
2 out of the 62 DECam ccds, N7 and S7, are not operational.

Photometry in the $F435W$, $F625W$, $F658N$ filters was collected with 
the Advanced Camera for Surveys (ACS) on board the Hubble Space Telescope (HST) 
for a region covering a FoV of $\approx 10\arcmin \times 9.5\arcmin$ centered on \omcen.
For more details about these observations see \citet[hereafter CS07]{castellani2007}.

DECam images were pre-reduced by using a pipeline developed by one of us, 
Photpipe\footnote{https://confluence.stsci.edu/display/photpipearmin/Photpipe}. 
Photpipe is a robust pipeline used by several time-domain surveys (e.g., SuperMACHO, ESSENCE, Pan-STARRS1; 
see \citealt{rest05, rest14}), designed to perform single-epoch image processing including image calibration.
We used Photpipe to perform bias subtraction, flat-fielding, cross-talk correction, geometrical distortion correction
and the image astrometric calibration.
 
Photometry was performed with a variant of DoPHOT \citep{schechter1993}, with a pipeline developed by one of us for 
reducing DECam images along the lines described in \citet{saha2010}.
DoPHOT uses an analytical function as a model point-spread function (PSF) for describing different object types. 
Aperture magnitudes are determined for bright isolated stars across each of the 60 operational chips of the camera.
The difference between the PSF and aperture magnitudes is mapped over the field of view,
accounting for chip to chip offsets, and this provides the correction factors for DoPHOT 
PSF magnitudes (see \citealt{saha2010}, for details). 
The same reduction pipeline performs a few quality selections on the catalog: 
i) stars that lie too close to the chip borders, i.e. less than 20 pixels away, are excluded;
ii) if a star has a close neighbor with significant comparative brightness is excluded; 
iii) stars with photometric errors more than 3 sigma the average photometric error for all objects in the 
corresponding 0.5 magnitude bin are excluded.

\section{Photometric calibration}\label{redu}

During the night of 2014 February 24, a Stripe82 field was observed in all the four filters, namely SDSS1048p0000.
We retrieved the photometry for the stars included in this field from the Sloan Digital Sky Survey (SDSS) and from the 
Pan-STARRS1 (PS1) catalogs. 
To transform the photometry into the DECam natural system we followed the approach of \citet{scolnic2015}.
We derived the transformations from PS1 and SDSS to the DECam system based on the photometry of 205 out 
of the 379 standard stars of the Next Generation Stellar Library (NGSL\footnote{https://archive.stsci.edu/prepds/stisngsl/}).
These standard stars span a color range $-1.0 \lesssim g-i \lesssim 2.5$ mag, and magnitudes 
are in the range $2 \lesssim g \lesssim 12$ mag (for more details on how these stars were selected see \citealt{scolnic2015}).
Fig.~2 shows the comparison of the SDSS and DECam $r$-band magnitudes of the selected standard stars versus
their SDSS $g-i$ color. To convert the magnitudes from the SDSS to the DECam system, and later from PS1 to DECam, 
we used an iterative process. We first selected a sample of stars by estimating the mean magnitude difference for each
0.15 mag color bin and kept only stars with a difference $\le$ 1 $\sigma$. The figure shows the selected (blue dots) and
the excluded (red) standard stars in the $r_{SDSS} - r_{DECam}$ vs $g_{SDSS} - i_{SDSS}$ plane. Another $1\sigma$ selection
was applied by using the fitted spline, and a new fit was performed. The final fitting spline is shown in the figure as
a green solid line. This spline was then used to transform the SDSS and PS1 photometry for 
the field SDSS1048p0000 into the DECam natural system.
Two different set of zero points (ZPs) were estimated by using the two photometries and they are 
listed in Table~3 together with their root mean square values (note
that the ZP for the $u$ filter was derived exclusively by using the SDSS photometry).
The ZPs derived by using the SDSS and the PS1 photometry agree at the 2\% level.

\begin{figure}
\label{fig1}
\includegraphics[height=0.4\textheight,width=0.5\textwidth]{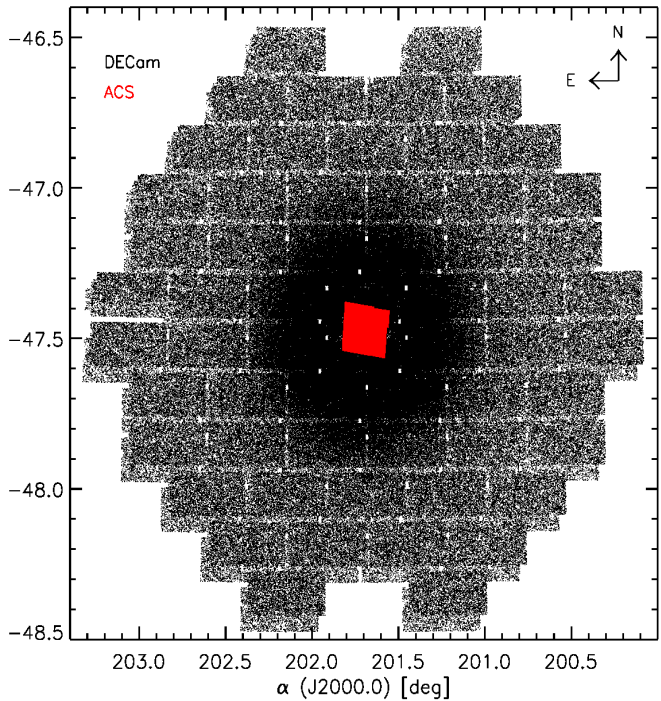} 
\caption{Field of view covered by DECam photometric catalog (black dots) and 
by the ACS catalog (red) across the globular cluster \omcen. The orientation is labeled in the figure. }
\end{figure}

\begin{figure} 
\label{fig2}
\includegraphics[height=0.4\textheight,width=0.5\textwidth]{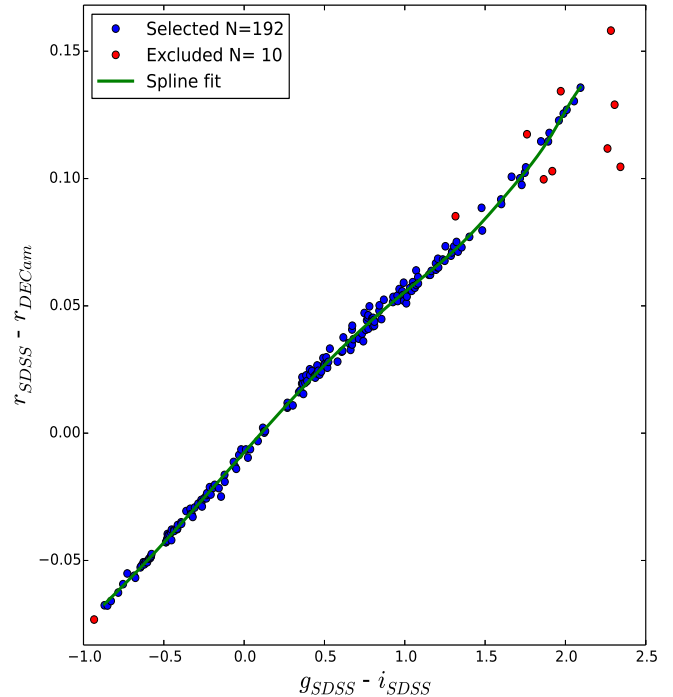}
\caption{Selected stars from the Next Generation Stellar Library in the $SDSS_r - DECam_r$ vs SDSS $g-i$ color plane. 
Sigma clipping selected (blue dots) and excluded (red) stars are shown. 
The spline fit is over-plotted as a green solid line.}     
\end{figure}

The instrumental magnitudes of the \omc catalog are corrected for aperture 
by selecting a few bright isolated stars for each DECam ccd. The aperture 
correction is estimated accounting for chip to chip offsets and mapping
the entire camera FoV. The best seeing image is selected as a reference for
each filter, and the photometry of the other images is rescaled to this one, 
after bringing all the exposures to 1 second.

We then calibrated the mean instrumental magnitudes by following the equation:

\begin{equation}
M_i = m_i  - K_i \times A_i + ZP_i
\end{equation}

where $M_i$ and $m_i$ are the calibrated and instrumental magnitudes, respectively, 
$K_i$ are the extinction coefficients for the different filters, 
$A_i$ the air masses of the reference observations and $ZP_i$ the zero points.
We used the extinction coefficients obtained by the DECam Legacy Survey (DECaLS)
collaboration \citep{li2016} for the $g,r,i$ filters, namely $K_g$ = 0.18, $K_r$ = 0.0875 and $K_i$ = 0.065. 
For the $u$ filter we used the value $K_u$ = 0.40 obtained from DECam 
multiple $u-$band observations of the Galactic bulge. 
The air masses of the reference observations are $A_u =$1.19, $A_g =$1.15, 
$A_r =$1.19 and $A_i = $1.19.  As ZPs we used the average of the ZPs derived 
by using the PS1 and SDSS photometry, namely 
$ZP_u = -$7.58, $ZP_g = -$5.48, 
$ZP_r = -$5.35 and $ZP_i = -$5.45.

The accuracy of the calibration is better than 2\% for the $gri$ filters, while is $\approx$ 5\% for the $u$ filter.

The calibrated photometric catalog for $\omc$ includes $686,746$ stars with a measurement in the $r$ and $i$ filters, 
$598,429$ with a measurement in the $g$ filter, and $398,604$ stars were also measured in the $u$ filter.
Fig.~3 shows the $i,\ u-i$ (left panel), $i,\ g-i$ (middle) and $i,\ r-i$ (right) color-magnitude diagrams (CMDs) for all stars 
observed in the FoV towards \omcen. The Signal to Noise ratio (S/N) is $\ge$ 20 down to $u \approx$ 23 mag, $g \approx$ 23 mag,
$r \approx$ 23 mag, and $i \approx$ 22.5 mag, which are the catalog limiting magnitudes. 
These CMDs are heavily contaminated by field stars, mostly thin and thick disk and halo stars (see the Galactic simulation for 
a 1 deg$^2$ FoV around $\omc$ of \citealt{marconi2014} and their Figure 11). 

The ACS photometry was kept in the VEGA system and we applied the camera 
charge transfer efficiency correction and the available ZPs 
for the $F435W$, $F625W$, $F658N$ filters following the prescriptions by 
\citet{sirianni2005}. For more details on the photometric calibration of 
this catalog see CS07.

\section{A clean sample of cluster stars}\label{clean}

To separate field and cluster stars we adopted a similar approach as suggested 
by \citet{dicecco2015}. To take advantage of the multi-band optical photometry 
available for globular clusters they estimated the cluster ridge lines using different CMDs based on the 
same magnitude ($r$) and different colors. 
To improve the precision of the cluster ridge lines the candidate cluster stars 
were selected according to their radial distance and to their photometric errors. 
Once the multiple ridge lines were estimated they generated a multi-dimensional 
CMD and the candidate cluster stars were selected using a variable $\sigma$-clipping 
over the entire magnitude range. This approach is seen to be quite robust, since 
they were able to separate candidate field and cluster stars for M~71,  a metal-rich 
globular projected onto the Galactic bulge.

However, the quoted method can be hardly adopted in a globular cluster like \omcen, 
due to the presence of well-defined multiple sequences mainly caused by a difference 
in metal content \citep{calamida2009, johnson2010}. Therefore, 
the separation was performed using a new improved approach. We estimated the ridge 
lines of the different sub-populations identified along the cluster red-giant branch (RGB),
the main sequence turn-off (MSTO) and the main sequence (MS). The horizontal branch (HB)
stars were not included since they are typically bluer than field stars. 
These ridge lines were estimated neglecting the stars located in the innermost cluster regions ($r\le$ 3\arcmin) 
and applying several cuts in radial distance and in photometric accuracy. 
Note that to fully exploit the current photometric catalog we only selected stars
 with accurate measurements in all the four $ugri$ bands.
 Once the ridge lines (seven) have been estimated we performed a linear interpolation 
 among them and generated a continuos multi-dimensional surface. Finally, we used 
 two different statistical parameters to separate field and cluster stars:

1) we estimated the cumulative standard deviation among the position of individual 
stars and the reference surface;

2) we associated a figure of merit to the distance in magnitude 
and colors among the individual stars and the reference surface.

We have performed a number of test and trials to sharpen the selection 
criteria to separate cluster and field stars. The approach was conservative, 
in the sense that we preferred to possibly lose some of the candidate 
cluster stars instead of including possible candidate field stars. 
A glance at Fig.~4 shows the advantages of the current approach. 
The left panel displays the color-color-magnitude diagram, $u-r$ vs $g-i$ vs $r$, for candidate field 
(gray dots)  and cluster (multi-color) stars seen from the front, while the right panel
shows the same plot but seen from the back.
Note that the candidate field stars display a smooth distribution both in magnitude and 
in color. In passing we note that the current approach can be applied to separate
field and cluster stars thanks to the opportunity 
to use the $u$ filter, since this band allows a better sensitivity to both 
effective temperature and metallicity. Fig.~5 shows the candidate field stars
in the $i,\ u-r$ CMD. No clear cluster sequence is present in the field
star sample in the entire magnitude range down to $i \approx$ 23 mag.
It is interesting to note the sequence of disk white dwarfs at -0.5$\le u - r \le$ 1.0 mag 
and $i \ge$ 18 mag  that were excluded from the cluster sample after applying our selection method. 


\begin{deluxetable}{lrc}
\tablecaption{Positional, photometric and structural parameters of the 
Galactic Globular Cluster \omc}\label{tbl-1}
\tablehead{
\colhead{Parameter}&
\colhead{         }&   
\colhead{Ref.\tablenotemark{a}} }
\startdata
$\alpha$ (J2000)                          &  201.694625   &     1    \\  
$\delta$ (J2000)                           &  -47.48330              &      1    \\  
$M_V$ (mag)\tablenotemark{a}              &    -10.3            &     3    \\  
$r_c$ (arcmin)\tablenotemark{b}             &  2.58               &     3	 \\  
$r_t$ (arcmin)\tablenotemark{c}              &  57.03               &     2	 \\  
$e$\tablenotemark{d}                      &  0.12            &     4	 \\  
$\sigma_V$ (km~s$^{-1}$)\tablenotemark{e} &  $17\pm1.6$     &     5    \\  
E(B-V)\tablenotemark{f}                          &  $0.11\pm0.02$  &     6    \\  
$(m-M)_0$ (mag)\tablenotemark{g}          &  $13.71\pm0.02\pm0.03$ &     1   \\  

\enddata 
\tablenotetext{a}{References: 1) Braga et al. (2016); 2) Harris (1996);
3) Trager, King \& Djorgovski (1995); 4) Geyer, Nelles \& Hopp (1983); 
5) Merrit, Meylan \& Mayor (1997); 6) Calamida et al.\ (2005)
$^a$ Total Visual magnitude.  
$^b$ Core radius. 
$^c$ Tidal radius.  
$^d$ Eccentricity. 
$^e$ Stellar central velocity, dispersion.  
$^f$ Reddening.  
$^g$ True distance modulus.}   
\end{deluxetable}


\begin{table*}
\caption{Log of the observations collected with DECam on the 4m Blanco telescope for \omc 
(CTIO, NOAO, proposal IDs: 2014A-0327, 2015A-0151, 2016A-0189, PIs: A. Calamida, A. Rest).}\label{table:2}
\begin{tabular}{l c c c c c }       
\hline\hline                
Name & Exposure time & Filter & RA & DEC & Seeing \\    
     &  (s)          &        & (hh:mm:ss.s) & (dd:mm:ss.s) & (arcsec) \\  
\hline                      
\hline
February 24, 2014\\
\hline
omegacen.u.ut140224.052814.fits  & 120 & u & 13:26:47.288 & -47:28:45.894 & 1.2 \\
omegacen.u.ut140224.053340.fits  & 120 & u & 13:27:00.889 & -47:33:26.593 & 1.3 \\
omegacen.u.ut140224.053907.fits  & 120 & u & 13:26:33.338 & -47:31:08.396 & 1.2 \\
omegacen.u.ut140224.054435.fits  & 120 & u & 13:26:13.607 & -47:34:28.294 & 1.6 \\
\hline
June 22, 2015 \\
\hline
omegacen.g.ut150622.035733.fits  & 250 & g & 13:26:47.047 & -47:28:45.995 & 1.2 \\
omegacen.g.ut150622.040213.fits  & 250 & g & 13:26:27.319 & -47:28:46.196 & 1.2 \\
omegacen.g.ut150622.040649.fits  & 250 & g & 13:26:27.337 & -47:32:06.295 & 1.2 \\
omegacen.g.ut150622.041130.fits  & 250 & g & 13:26:47.058 & -47:32:06.194 &  1.1 \\
\hline
March 4, 2016 \\
\hline
omegacen.r.ut160304.072010.fits  & 80 & r & 13:26:48.138 & -47:27:41.994 & 0.8 \\
omegacen.r.ut160304.072202.fits  & 80 & r & 13:26:40.258 & -47:27:41.695 & 0.8 \\
omegacen.r.ut160304.072350.fits  & 80 & r & 13:26:40.229 & -47:26:21.494 & 0.8 \\
omegacen.r.ut160304.072538.fits  & 80 & r & 13:26:48.178 & -47:26:21.793 & 0.8 \\
\hline
\hline
\end{tabular}
\begin{tablenotes}
\small
\item This table is available in its entirety in a machine-readable form in the online journal.
 \end{tablenotes}
\end{table*}


\begin{table*}
\caption{Zero points derived for Stripe 82 field SDSS1048p0000 by using SDSS and PS1 photometry.}\label{table:3}     
\begin{tabular}{l c c c c c c c c}       
\hline\hline                
System &  $u$ & $g$ & $r$ & $i$ & $rms_u$ & $rms_g$ & $rms_r$ & $rms_i$ \\    
\hline                      
\hline
PS1   & \ldots  &  -5.466$\pm$0.002 & -5.350$\pm$0.002 & -5.440$\pm$0.002 & \ldots & 0.03 & 0.02 & 0.03 \\
\hline                      
SDSS & - 7.577$\pm$0.007 & -5.495$\pm$0.003 & -5.352$\pm$0.003 & -5.462$\pm$0.002 & 0.05 & 0.04 & 0.03 & 0.03 \\
\hline
\hline
\end{tabular}
\end{table*}


\begin{table}
\caption{Values of the peaks and Full-Width half maximum (FW) for the three Gaussians that fit the three sub-populations along
\omc MS: P$_1$ for the red MS, P$_2$ for the MS-a and P$_3$ for the blue MS. See text for more details.}\label{table:3}     
\begin{tabular}{c c c c c c c}       
\hline\hline                
$\Delta$ Mag & P$_1$ & FW$_1$ & P$_2$ & FW$_2$ & P$_3$  & FW$_3$   \\    
\hline                      
\hline
5' $\le r <$ 10' \\
\hline
18    $\le i <$ 18.5 &  0.00  &   0.07  &  0.04  &  0.16  & \ldots  &  \ldots   \\
18.5 $\le i <$ 19    &  0.00  &   0.09  &  0.04  &  0.21  & \ldots  &  \ldots   \\
19    $\le i <$ 19.5 &  0.00  &   0.12  &  0.04  &  0.26 &  \ldots  &  \ldots  \\
19.5 $\le i <$ 20    &  0.00  &   0.12  &  0.03  &  0.30 &  -0.10  &  0.05  \\
20    $\le i <$ 20.5 &  0.00  &   0.19  &  0.19  &  0.19 &  -0.10  &  0.26  \\
20.5 $\le i <$ 21    &  0.01  &   0. 21 &  0.20  &  0.09 &  -0.18  &  0.16  \\
\hline                      
10' $\le r <$ 15' \\
\hline          
18    $\le i <$ 18.5 & 0.00  &   0.04 &  0.03  &   0.11 &   \ldots   &   \ldots   \\
18.5 $\le i <$ 19    & 0.00  &   0.05 &  0.02  &   0.13 &   \ldots   &  \ldots   \\
19    $\le i <$ 19.5 & 0.00  &   0.06 &  0.03  &   0.19 & -0.07   &   0.03  \\
19.5 $\le i <$ 20    & 0.00  &   0.09 &  0.01  &   0.20 & -0.09   &   0.06   \\
20    $\le i <$ 20.5 & 0.00  &   0.09 &  0.01  &   0.28 & -0.12   &   0.07   \\
20.5 $\le i <$ 21    & 0.00  &   0.09 & -0.02  &   0.27 & -0.15   &   0.07   \\
\hline
15' $\le r <$ 66' \\
\hline                      
18    $\le i <$ 18.5 &  0.00 &  0.05 & 0.06  &  0.07  &   \ldots   &   \ldots   \\
18.5 $\le i <$ 19    &  0.00 &  0.07 & 0.06  &  0.16  &   \ldots   &   \ldots   \\
19    $\le i <$ 19.5 &  0.00 &  0.07 & 0.08  &  0.09  &  -0.07  &  0.05  \\
19.5 $\le i <$ 20    &  0.00 &  0.05 & 0.03  &  0.09  &  -0.08  &  0.09  \\
20    $\le i <$ 20.5 &  0.00 &  0.07 & 0.04  &  0.16  &  -0.11  &  0.09   \\
20.5 $\le i <$ 21    &  0.00 &  0.09 & 0.06  &  0.16  &  -0.13  &  0.14  \\
\hline
\hline
\end{tabular}
\end{table}


\section{The cluster color-magnitude diagrams}
Following the procedure described in section \S 4 we ended up with a catalog of 
266,769 cluster members with a measurement in all the four $ugri$ filters. The $u$-band photometry is limiting
the depth of the catalog, but it was essential in allowing the cluster and field star separation.
Fig.~6 shows the same CMDs of Fig.~3 but for cluster members only. No selection in photometric accuracy is applied.
All the cluster sequences are well-defined, including the extreme horizontal branch (EHB) at 
$u-i \approx$ -1,  $g-i \approx$ -0.7, $r-i \approx$ -0.3 and 18 $< i <$ 20 mag, and the 
white dwarf (WD) cooling sequence at $u-i \approx$ -1,  $g-i \approx$ -0.7, $r-i \approx$ -0.3 and $i <$ 21 mag. 
The CMDs reach $i \approx$ 21.5 mag with $S/N >$ 50. 

To verify the plausibility of the photometric calibration we used theoretical isochrones and 
Zero Age Horizontal Branch (ZAHB) loci from the BASTI database\footnote{http://albione.oa-teramo.inaf.it}.
These models are in the Sloan photometric system \citep{fukugita1996} and were transformed to 
the DECam system by using the empirical transformations derived in section \S 3.
Extinction coefficients in the $ugri$ filters were estimated by using the \citet{cardelli89} reddening law and 
DECam filter transmission functions. We obtained $A_u$ = 1.70$\times A_V$, $A_g$ = 1.18$\times A_V$, 
$A_r$ = 0.84$\times A_V$, and $A_i$ = 0.63$\times A_V$.
We used an absolute distance modulus of $\mu_0$ = 13.71$\pm$0.02$\pm$0.03 mag \citep{braga2016} and 
a reddening of $E(B-V)$ = 0.11 $\pm$0.02 mag \citep{calamida2005}.
We selected two isochrones for the same age, $t$ = 12 Gyr, and different metallicities, 
namely $Z$ = 0.004, $Y$ = 0.251, and $Z$ = 0.0006, $Y$ = 0.246. These values,
 -1.84 $\le [Fe/H] \le$ -1.01, approximately bracket 
the bulk of \omc metallicity dispersion \citep{calamida2009}.
The agreement between theory and observations is quite good in the entire magnitude range in 
all the three CMDs. The HB in the $i,\ u-i$ CMD is slightly bluer than the ZAHBs. This effect might be due to the calibration 
uncertainties, $\approx$ 5\%, and to uncertainties in the $u$-band bolometric correction of the models.

\begin{figure*}
\begin{center}
\label{fig3}
\includegraphics[height=0.75\textheight,width=0.75\textwidth, angle=90]{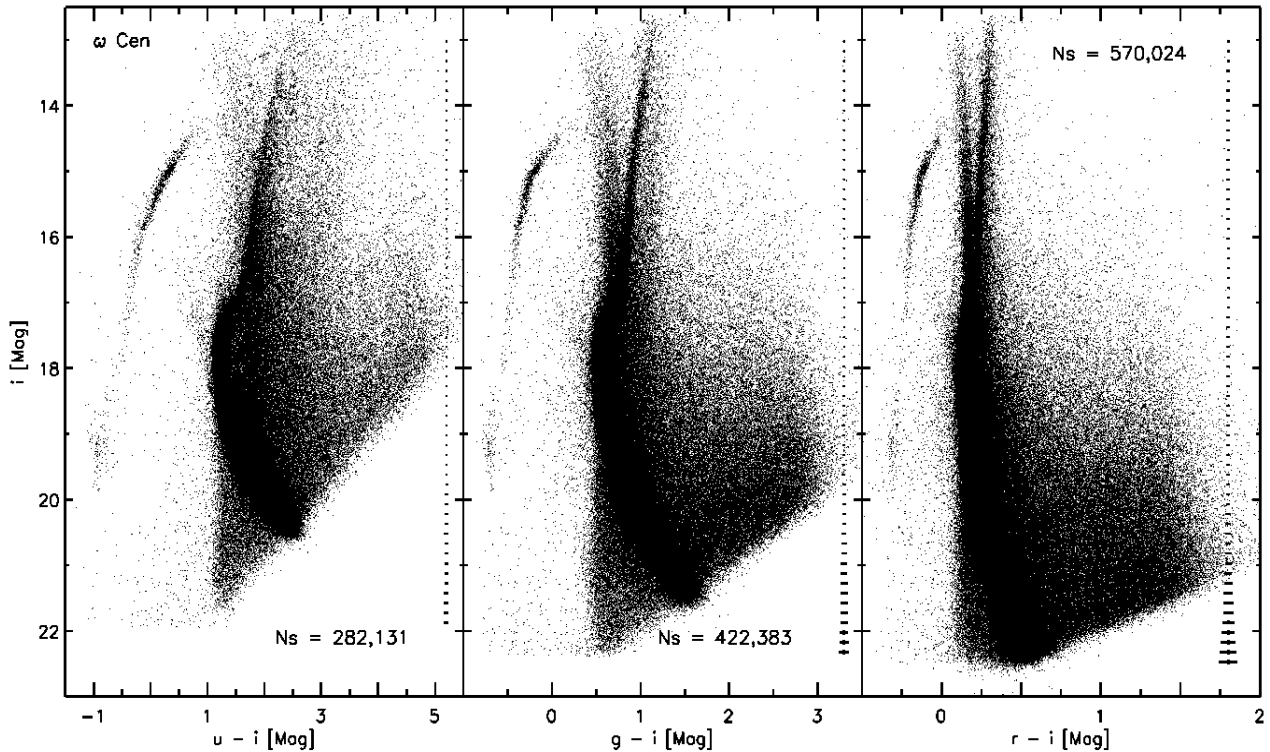} 
\caption{DECam {\it ugri} color-magnitude diagrams towards \omcen. Error bars are marked.}
\end{center}
\end{figure*}

\section{Astrometry and coordinate system}
The astrometric calibration of \omc DECam catalog to the equatorial system J2000 
was performed by using Photpipe and the Two Micron All Sky Survey \citep{cutri2003}
catalog of stars as a reference. The final accuracy is better than 0.03" in both right ascension and
declination.

The astrometry of the ACS catalog was performed by matching the photometry with stars
from the catalog of \citet{vanLeeuwen2000} with proper motions and membership probabilities 
(for more details see CS07).

We matched the ACS and DECam photometric catalogs for cluster members by using a 0.5" searching radius. 
The matched catalog includes 1,722,810 stars covering a FoV of 2.3$\times$2.2$\deg$ across 
\omc and including photometry in seven photometric bands, namely $F475W, F625W, F658N, u, g, r, i$ (see Fig.~1).
To our knowledge, this is the largest multi-band data set ever collected for a Galactic 
globular cluster after our ACS-WFI catalog published in CS07.
 
The equatorial coordinates $\alpha$ and $\delta$ in degrees were converted to cartesian coordinates by 
following the prescriptions of \citet{vandeven2006} with the cluster center at $\alpha_0$ =  201.694625\deg
and $\delta_0$ = -47.48330\deg \citep{braga2016}. Setting $x$ in the direction of West and $y$ 
in the direction of North:\newline

$x = -r_{0}\cdot cos\delta \cdot sin\Delta \alpha$
\newline

$y = r_{0}\cdot (sin\delta \cdot cos\delta _{0} - cos\delta \cdot sin\delta _{0}\cdot cos\Delta \alpha )$
\newline

where $\Delta \alpha = \alpha - \alpha_0$ and $r_0 = 10,800/\pi$ to have $x$ and $y$ in arcminutes. 

We then projected the cartesian coordinates $x$ and $y$ with the
$x$ and $y$ axes aligned with the observed major and minor axes of \omcen, respectively.
To accomplish this we rotated the coordinates by the position angle of the cluster, defined as the 
angle between the major axis and the North direction measured counterclockwise
by using a value of 100$\deg$ \citep{vandeven2006}.
The combined ACS-DECam photometric catalog with coordinates aligned on 
\omc major and minor axis will allow us to investigate the behavior of the cluster 
different sub-populations as a function of distance from the center.

\section{The main-sequence split}
Based on HST photometry \citet{anderson2002} before, and later \citet{bedin2004}, revealed that \omc MS is bifurcating
into two main components, the so called blue-MS (bMS) and the red-MS (rMS). 
The color difference between the two sequences changes with magnitude, and they are clearly separated
in the magnitude interval 20.5 $< V <$ 22. The HST observations included a central field and a field 
located at $\approx$ 17\arcmin~ from the cluster center. A spectroscopic follow-up by \citet{piotto2005} 
found that bMS stars are more metal-rich than rMS stars counter to expectations.
These authors proposed that bMS stars constitute a
helium-enhanced sub-population in the cluster to explain the observed anomaly.
\omc MS split was also found by \citet{sollima2007a} based on VLT photometry.
They showed that the two sequences are still well-separated at $\approx$ 26\arcmin~ from 
the cluster center and that bMS stars are more centrally concentrated compared to rMS stars.
The ratio of bMS and rMS stars decreases from a value of $\approx$ 0.28 
at a distance from the cluster center of $\approx$ 7\arcmin~ down to 0.15 for distances larger 
than 19\arcmin. \citet{bellini2010}, by using deep HST observations collected in different 
filters from the ultraviolet to the red, showed the presence of a third MS, named MS-a, 
that better separates in the $F435W,\ F336W - F435W$ CMD, and seems to be connected 
to \omc faintest sub-giant branch (SGB), the so called SGB-a \citep{ferraro2004}, and the 
reddest and most metal-rich RGB, the so-called RGB-a \citet{pancino2000}, and named $\omega$3 by us (CS07). 
However, these analyses were limited to the cluster central region (HST data) and up to a distance 
of 25\arcmin~ (HST and VLT data). DECam photometry and the ability to remove the field component by using 
color-color-magnitude diagrams opens up the possibility to investigate the spatial distribution of 
the two MSs until \omc tidal radius. 

The left panel of Fig.~7 shows a zoom of DECam $i,\ g-i$ CMD in the magnitude interval 18.0 $\le i \le$ 21.0. 
The split of the MS is evident, with the bMS well-separated from the rMS in the magnitude range 19.0 $\le i \le$ 21.0.  
This is the first time that \omc MS split is observed with a 4m-class ground-based telescope. 
The $g-i$ color distance between the two sequences is changing 
with magnitude and reaches a maximum of $\approx$ -0.18 mag at $i \approx$ 21 mag. Unfortunately, our catalog
does not have the sufficient photometric accuracy to allow us to investigate the behavior of the MS at fainter magnitudes.
The ridge line of the rMS is over-plotted on the $i,\ g-i$ CMD of Fig.~7 as a red solid line. 
The ridge line color at the corresponding magnitude was subtracted to each star observed color 
to straighten the MS. The result of this process is shown in the right panel of the figure. 
We then estimated the distance of each star from the cluster center by using the coordinates aligned with 
the major and minor axes and divided the stars in three concentric annuli from 5 to 66\arcmin, 
including approximately the same number of stars per radial annulus.
Stars were then divided in six 0.5 magnitude bins from $i$ = 18 down to $i$ = 21. 
Fig.~8 shows the star $g-i$ observed color minus the ridge line color histograms for the six magnitude intervals 
for the annulus in the distance interval 10 $< r <$ 15\arcmin. 
The panels show that the color distributions are asymmetric, 
being skewed towards the red in the entire magnitude range, and they
separate in two main peaks starting at $i \approx$ 19 mag. 
The skewness is probably due to the presence of the third MS (MS-a), that the
accuracy of the photometry and the $g-i$ color sensitivity does not allow us to separate 
it from the rMS, and to the presence of blends and unresolved binaries.
We fitted the six histograms with three Gaussians reproducing the rMS ($P_1$), the MS-a ($P_2$) and the
bMS ($P_3$), respectively. The three Gaussian functions used in the fit and their sum are shown in 
the figure as red (rMS), green (MS-a), blue (bMS) and black solid lines. The peaks and the Full-Width Half Maximum 
values of  the Gaussians are indicated in the plots and listed in Table~4.

DECam photometry clearly shows that \omc MS split is present at all distances from the cluster center
until the tidal radius. 
The $g-i$ color separation between the rMS ($P_1$) and the bMS ($P_3$) is the same, within the uncertainties, 
for the three different annuli, going from $0.07$ to $0.15$ mag for 10 $\le r <$15\arcmin, according to the magnitude interval.

\subsection{The ratio of blue and red main-sequence stars}

To characterize the spatial distribution of  \omc MS stars we computed the ratio of
bMS and rMS stars, $r(bMS/rMS) = N(bMS)/N(rMS)$, as a function of the radial distance.
To select the sample of bMS and rMS stars we first produced
a 3D CMD for stars in the magnitude interval 19.25 $< i <$ 20.5,
where the MS best separates. 
Fig.~9 shows the 3D CMD: color $g-i$, magnitude $i$, and luminosity function.
To overcome subtle problems in constraining the position of the MS peaks 
caused by the binning of the data, we associated to each star a 
Gaussian kernel with a sigma equal to its intrinsic error in the $g-i$
color measurement. The green surface was computed by summing all the 
individual Gaussians over the entire color and magnitude range. 
A glance at the surface discloses two distinct backbones tracing the 
bMS and the rMS (blue and red solid lines, respectively).
To further improve the identification of bMS and 
rMS stars, the blue and the red lines display the peaks of the two sequences, 
while the black solid line marks the valley between the two different 
sub-populations, i.e. the relative minimum between the two relative maxima.

The 3D CMD plotted in Fig.~9 clearly shows that the separation 
between bMS and rMS stars is far from being trivial, since the difference 
in color is magnitude dependent. Moreover, the MSs associated to the 
two sub-populations display, at fixed magnitude, different broadenings in color. 
To overcome thorny problems in the selection criteria adopted to identify bMS and 
rMS stars, we adopted an incremental approach. Firstly, we only 
selected stars that are located within $\pm$ 0.02 mag from the blue and the 
red backbone in the 19.25 $< i <$ 20.5 magnitude interval.
This means we selected stars with a distance in $g-i$ color 
from the backbone of $\Delta =$ 0.04 mag. The adopted minimum color bin 
was driven by the typical color uncertainty in the selected magnitude range. 
We then repeated the same selection but increasing the distance in color from 
the backbone. Note that the bMS and the rMS samples never overlap, 
since the inner boundary is traced by the valley plotted 
in Fig.~9. To improve the statistics of the two samples, the range in color was 
increased up to 0.3 mag. We performed a number of tests and this color bin
is a good compromise between the width in color of the bMS plus the rMS and the 
contamination of field stars. 
It is clear that with a wider bin in color, we mainly select stars that 
are located along the slopes either of the bMS or of the rMS backbone. 
As a whole we ended up with 28 bins in $g-i$ color ranging 
from 0.02 to 0.3 mag. 

To validate the criteria adopted for the selection of bMS and rMS stars 
Fig.~10 shows the quoted selection in the $i,\ g - i$ CMD. 
The left panel shows the selection based on a $g-i$ color range
$\Delta =$ 0.04 mag. Note that for this selection the blue and the 
red samples are approaching the valley in the bright regime, $i <$ 19.4 mag,
but they are well separated in the faint regime. The middle panel shows an 
intermediate selection in which the two sub-populations already approached the valley 
over the entire magnitude range, while the right panel shows bMS and rMS stars
selected by using a wider $g-i$ color range.

Fig.~11 shows the 3D plot of the ratio between bMS and rMS stars 
as a function of the radial distance and of the $g-i$ color range 
used in the selection for the entire sample of stars 
included in the 19.25 $< i <$ 20.5 magnitude interval (left panel), for 
only the candidate cluster stars (middle), and for only the candidate 
field stars (right). The different color selections were plotted with different 
arbitrary colors to highlight the difference when moving from narrower 
to wider color bins. Table~5 lists the number of bMS and rMS stars and their ratio
for the samples of candidate cluster and field stars for $\Delta =$ 0.15 mag.
The ratios plotted in the three different panels and listed in the table
display several relevant features worth being discussed in more detail:

1) The ratio between bMS and rMS is far from being constant across the body 
of the cluster. It shows a well defined minimum, $r(bMS/rMS) = 0.17\pm0.005$, 
for radial distances of 20 $< r <$ 25\arcmin, in which the ratio decreases by almost a factor of two
from the half-mass radius (5\arcmin), and then it starts to steadily increase. 
Note that all previous investigations concerning the radial trend of bMS and 
rMS stars reached a maximum distance of $\approx$ 25\arcmin~ from the cluster center.

2) The population ratios display two relative maxima in approaching the cluster center 
and for radial distances of the order of 45--50\arcmin. 
These findings are independent of the $g-i$ color bin used to select the sample
of bMS and rMS stars and indeed the radial trends are quite similar
when moving from the narrower to the wider bin. 
Moreover, the current finding is also independent of the approach used 
to select candidate cluster and field stars. The population ratios are 
similar in the left panel of Fig.~11, where the ratio is the entire sample 
of stars, and in the middle panel, where it is based on only 
candidate cluster members. 

3) Data plotted in the middle panel of Fig.~11 further support the evidence 
that the maximum in the population ratio is attained in the outskirts of 
\omc  ($r \approx$ 48\arcmin), where bMS stars are overwhelming rMS stars, 
the ratio being of the order of 1.2 (see Table~5).
The population ratio increase in the innermost cluster regions only produces a 
relative maximum. This trend becomes, for statistical reasons, more clear when moving 
from the narrower to the wider color bins.

4) The radial trends of the population ratio display a steady decrease when moving 
from the maximum ($r \approx$ 48\arcmin) to the truncation radius ($r \approx$ 57\arcmin) 
of the cluster. This  decrease is affected by statistics, and indeed, the narrower color bins are slightly 
noisier when compared to the wider ones.

The population ratios based on only candidate field stars plotted in 
the right panel of Fig.~11 show a trend that at glance might appear counterintuitive, 
since they display a well defined maximum in approaching the innermost cluster 
regions, with $r(bMS/rMS) = 4.04\pm0.15$. The expected trend would have been 
a relatively flat distribution, as observed for distances larger than $\approx$ 35\arcmin.
However, this is a consequence of the fact that we are plotting the ratio 
between bMS and rMS stars and not the star counts. The ratios 
attain similar values, but the number of bMS and rMS stars among candidate cluster 
and field stars are significantly different. 
In passing we also note that the ripples showed by the cumulative population 
ratios are a consequence of small fluctuations in the number of blue and 
red candidate field stars.  

To further quantify the difference in star counts between cluster and field candidate 
stars the left panel of Fig.~12 shows the ratio of blue candidate 
field and cluster stars. The radial trends show that star counts of blue field 
stars are at least two order of magnitude smaller than those of blue cluster stars. 
This means that the maximum in the bMS and rMS star ratio in the right panel of
Fig.~11 is caused by a non-perfect separation between candidate field and 
cluster stars. However, the star counts of blue field stars are at most 
a few hundredths of the candidate blue cluster members. This means that they do 
not affect the current findings concerning the radial trend of the population ratio.  
The same outcome applies to the minimum and to the maximum of the population ratios. 
The candidate cluster bMS stars outnumber the candidate blue field stars up to 
radial distances of the order of 35\arcmin. In these regions the two samples attain, 
within the errors, star counts $log (N_{field}/N_{MS}) \approx$ 0. At larger radial distances the 
candidate field stars outnumber, as expected, candidate cluster members. The ratios 
plotted in the right panel of Fig.~12 are based on candidate red field stars 
and candidate rMS stars. The radial trends are similar to the trends of the blue stars. 

Data plotted in Fig.~12 are suggesting that we are quite confident concerning the 
robustness of the population ratios for radial distances smaller than 
$\approx$ 50\arcmin. The decrease in the population ratio at larger distances 
requires independent confirmation possibly based on photometric catalogs 
selected using proper motion.

\begin{figure*}
\centering
\begin{minipage} [l]{0.48\textwidth}
\centering
\includegraphics[height=8.5cm,width=8.5cm]{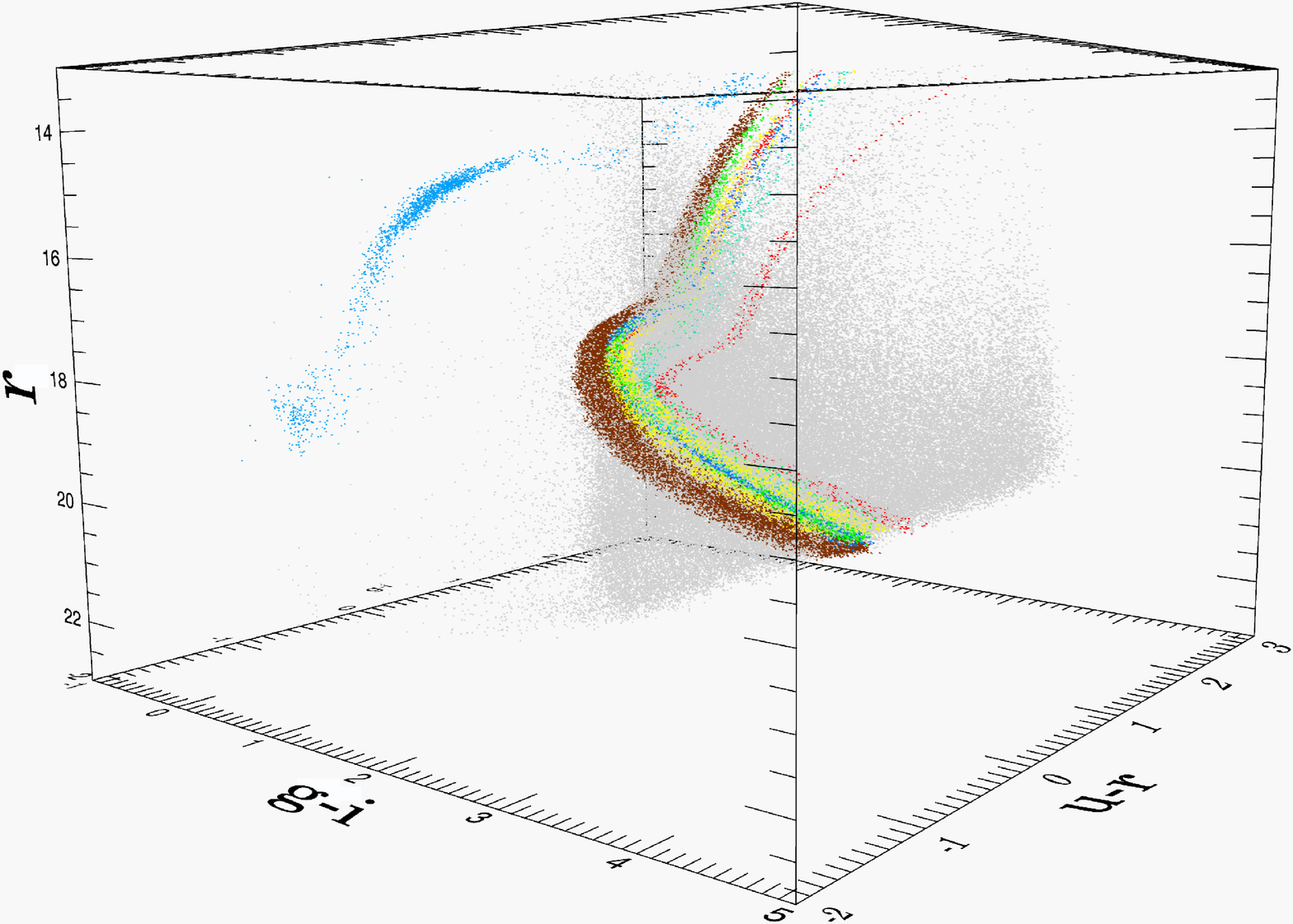}
\end{minipage}
\begin{minipage}[r]{0.48\textwidth}
\centering
 \includegraphics[height=8.5cm,width=8.5cm]{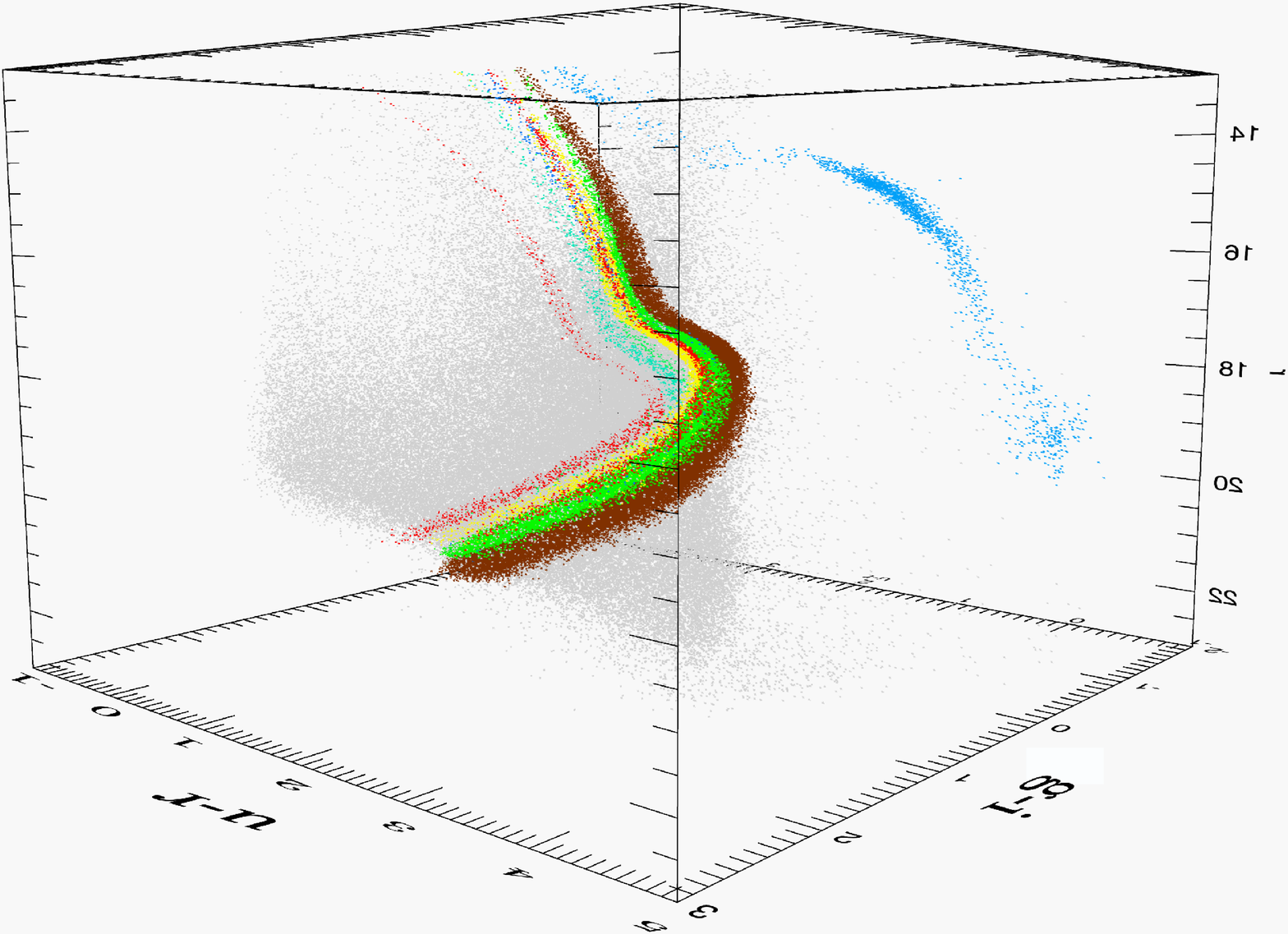}
\end{minipage}
\caption{\label{fig4}{\it ugri} DECam color-color-magnitude diagram of \omc members (different colors mark
different selected cluster sequences) and field stars (gray dots) seen from the front (top panel) 
and the back (bottom).}
\end{figure*}

\begin{figure}
\begin{center}
\label{fig5}
\includegraphics[height=0.40\textheight,width=0.5\textwidth]{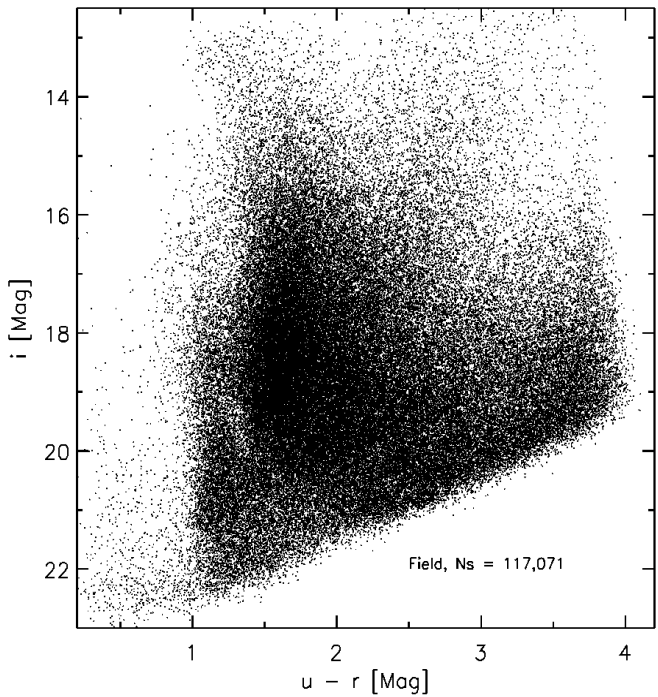} 
\caption{DECam $i,\ u-r$ color-magnitude diagram of candidate field stars in the observed field of view.}
\end{center}
\end{figure}

\begin{figure*}
\begin{center}
\label{fig6}
\includegraphics[height=0.75\textheight,width=0.7\textwidth, angle=90]{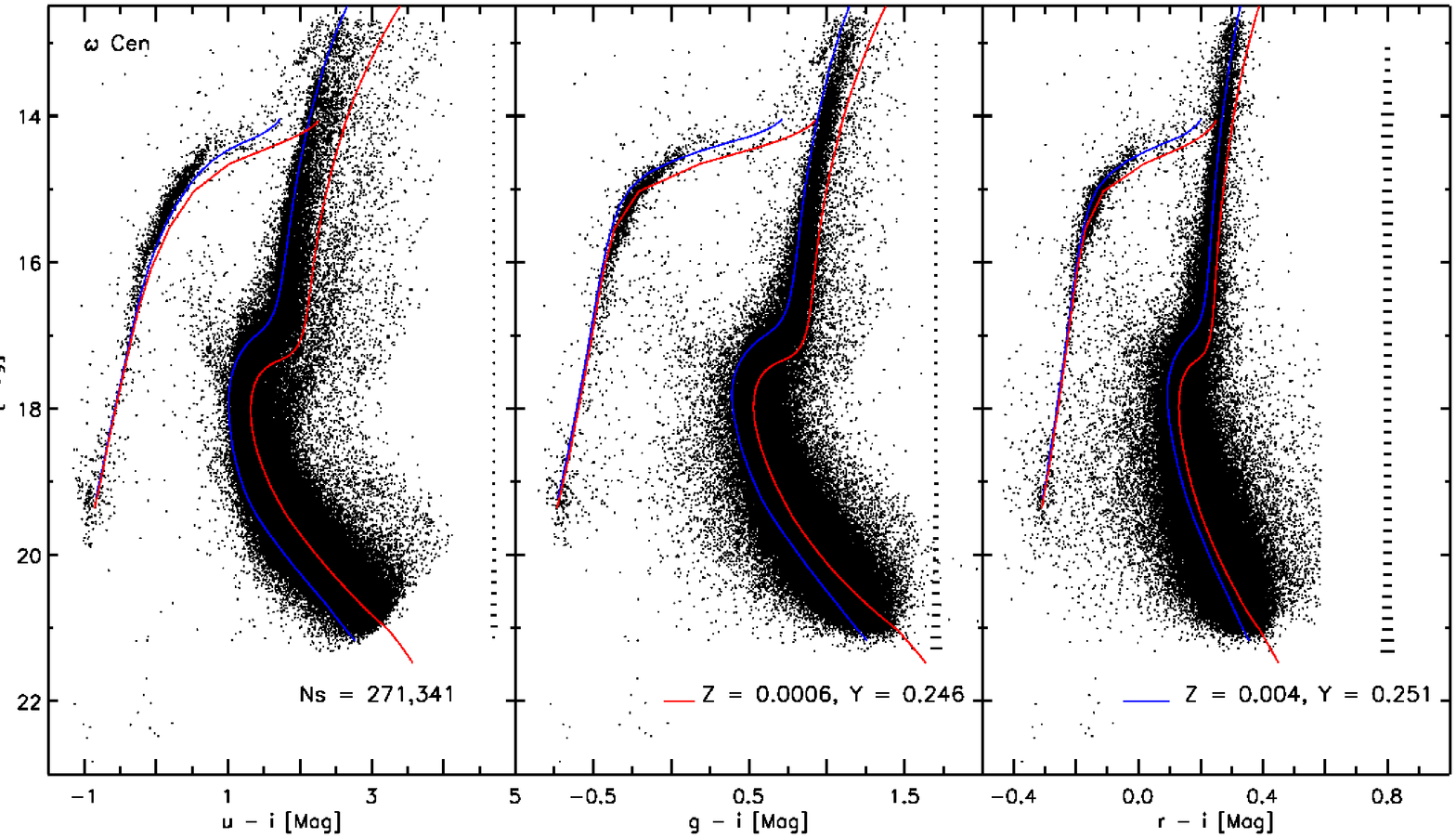} 
\caption{DECam {\it ugri} color-magnitude diagrams of \omc cluster members. Isochrones
for the same age, $t$ = 12 Gyr, and different metallicities are over-plotted (see labeled values). 
The respective zero age horizontal branch (ZAHB) tracks are also shown. Error bars are marked.}
\end{center}
\end{figure*}

To further support our result we plotted in Fig.~13 the $i,\ g-i$ CMDs
of candidate cluster (left panels) and field (right) stars for two external 
radial annuli, 30 $\le r <$ 35 and 35 $\le r <$ 60\arcmin, respectively. 
The figure shows that our method to separate cluster and 
field stars is very effective until large distances, $r \ge$ 30\arcmin, from the cluster center.
However, as discussed before and illustrated in the previous figures, 
some \omc stars might be miss-classified as field stars in the more
internal regions of the cluster. 
A residual contamination of field stars in the cluster MS samples might also be present 
at large distances from the center (see bottom left panel).

\subsection{Star counts across the body of the cluster}

\begin{figure*}
\begin{center}
\label{fig7}
\includegraphics[height=0.75\textheight,width=0.7\textwidth, angle=90]{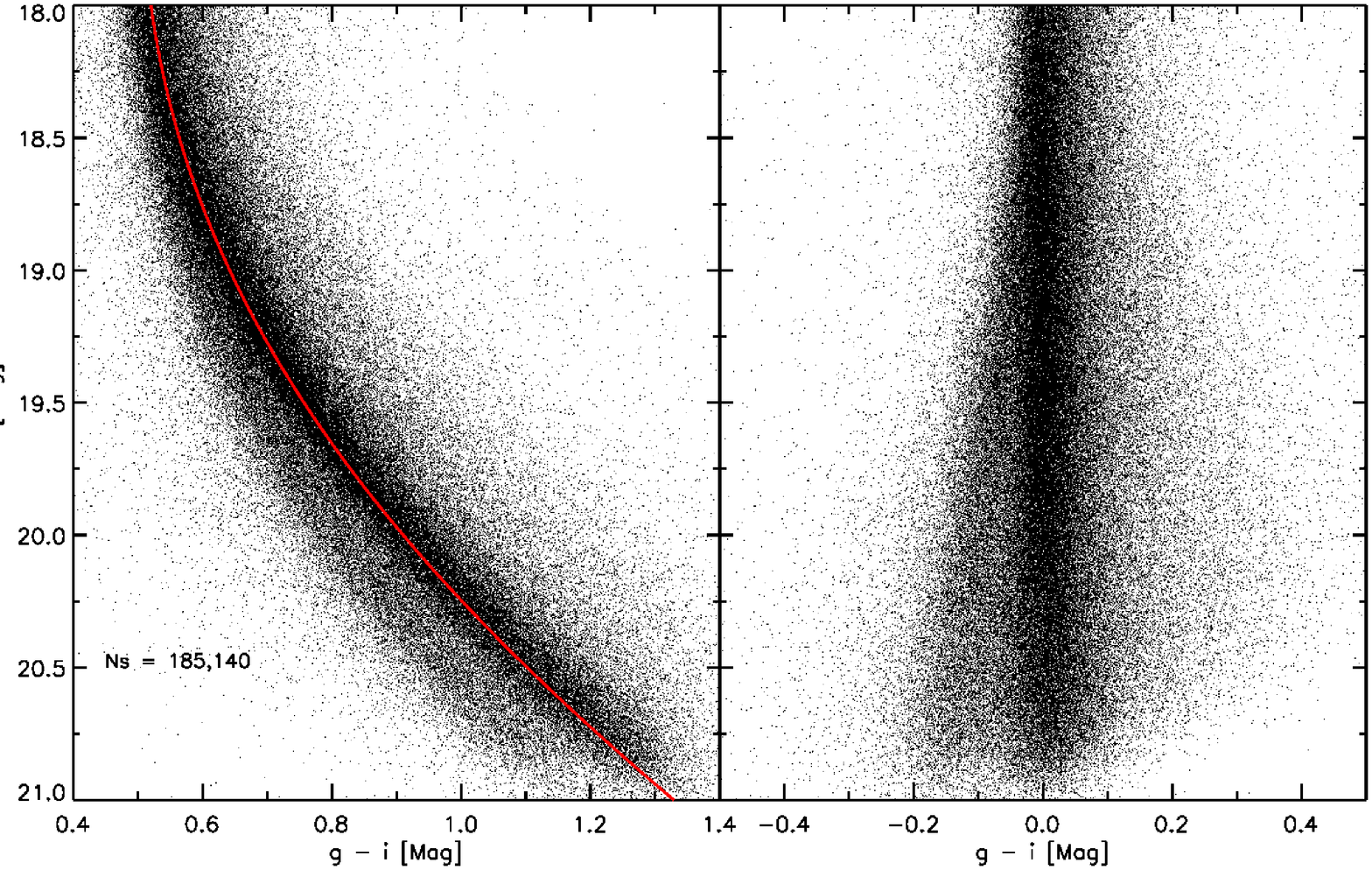} 
\caption{Left - Zoom of the \omc $i,\ g-i$ color-magnitude diagram. The ridge line for the rMS is 
over-plotted as a red solid line. 
Error bars are marked.
Right - Observed star $g-i$ color minus the ridge line color for the respective $i$ magnitude. }
\end{center}
\end{figure*}

\begin{figure}
\begin{center}
\label{fig8}
\includegraphics[height=0.75\textheight,width=0.5\textwidth]{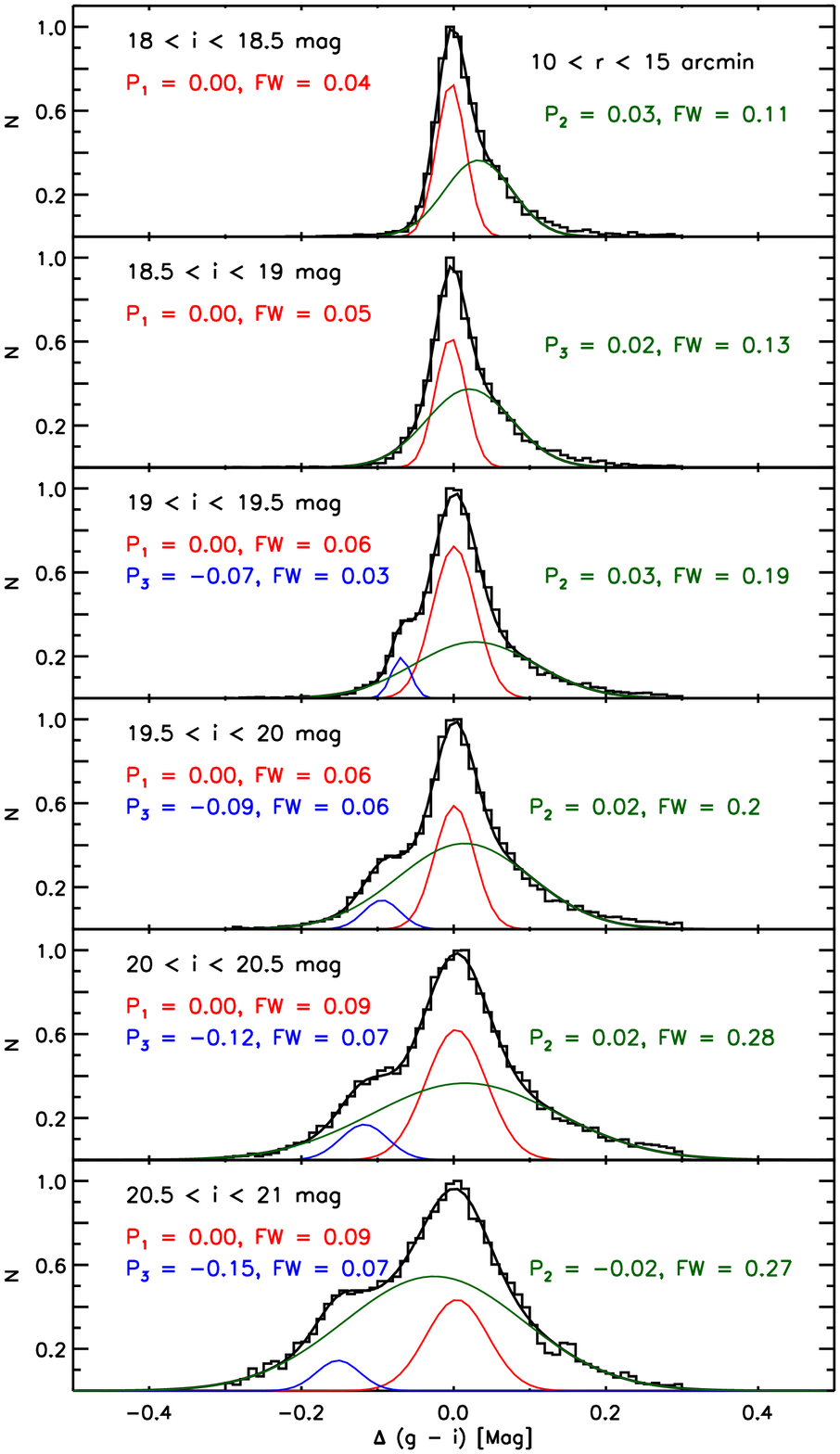} 
\caption{$g - i$ color histograms of \omc straightened main-sequence for six $i$-band magnitude intervals 
for the distance annulus 10 $< r <$ 15\arcmin. The three Gaussian functions and their sum used to fit the histograms
are indicated as red (rMS), green (MS-a), blue (bMS) and black (total) solid lines.
The Peak (P) and Full-Width Half Maximum (FW) values for the Gaussians are shown.}
\end{center}
\end{figure}

The spatial distribution of \omc MS stars appears to be even more 
complicated than suggested by the radial gradient. The density map of the 
ratio between bMS and rMS stars is plotted in the left panel of Fig.~14. 
The population ratio shows a clumpy distribution, 
with a well-defined North/South asymmetry in the outermost cluster regions, 
being the bMS stars significantly more abundant in the Northern quadrants. 
Thus suggesting that the clumping in the radial gradient might be 
associated to azimuthal variations across the body of the cluster. 
It is worth noting that the main over-density of bMS stars is 
pointing towards the Galactic center (GC) \citep[see the arrows plotted 
in the left panel of Fig.~14]{dauphole1996,leon2000}.
These findings seem to suggest a connection between the spatial 
distribution of bMS stars and \omc dynamical evolution. A more quantitative 
analysis of the difference between bMS and rMS stars in these cluster regions 
does require new kinematic and spectroscopic data.

The anonymous referee suggested us to investigate whether possible 
extinction variations across the body of the cluster could affect the 
current population ratio. To verify that our result is not affected by 
foreground reddening, we downloaded reddening values provided by 
\citet{schlafly2011}\footnote{http://irsa.ipac.caltech.edu/applications/DUST/}
for the region covered by the current photometric catalog. 
The reddening color density map of the observed field is 
shown in the right panel of Fig.~14. The reddening in the current FoV 
has a minimum value of $\approx$ 0.07 mag and a maximum of $\approx$ 0.18, with 
a mean reddening  of $E(B-V) =$ 0.11 mag and a total dispersion of $\sigma_{E(B-V)}$= 0.02 mag. 
These values are in very good agreement with the dispersion $\sigma_{E(B-V)}  \lesssim$ 0.03 mag 
found by \citet{cannonstobie1973} and  \citet{calamida2005} based on photometric studies 
of \omcen. This low differential reddening could move stars from the 
blue to the red MS sample. However, the observed over-densities of bMS stars 
when moving towards the outermost cluster regions cannot be caused by an increase 
in the extinction. An increase in differential reddening causes a decrease in 
the population ratio, since truly bMS stars are moved into the rMS sample. 
Therefore, the current population ratio can be considered as a lower limit 
to the real one.

On the other hand, the presence of higher dust extinction ($E(B-V) \approx$ 0.15--0.16 mag)
in the South-West corner of the cluster might explain the observed increase in rMS 
stars in the cluster regions located along the direction of \omc proper motion.

\begin{figure}
\begin{center}
\label{fig9}
\includegraphics[height=0.4\textheight,width=0.5\textwidth]{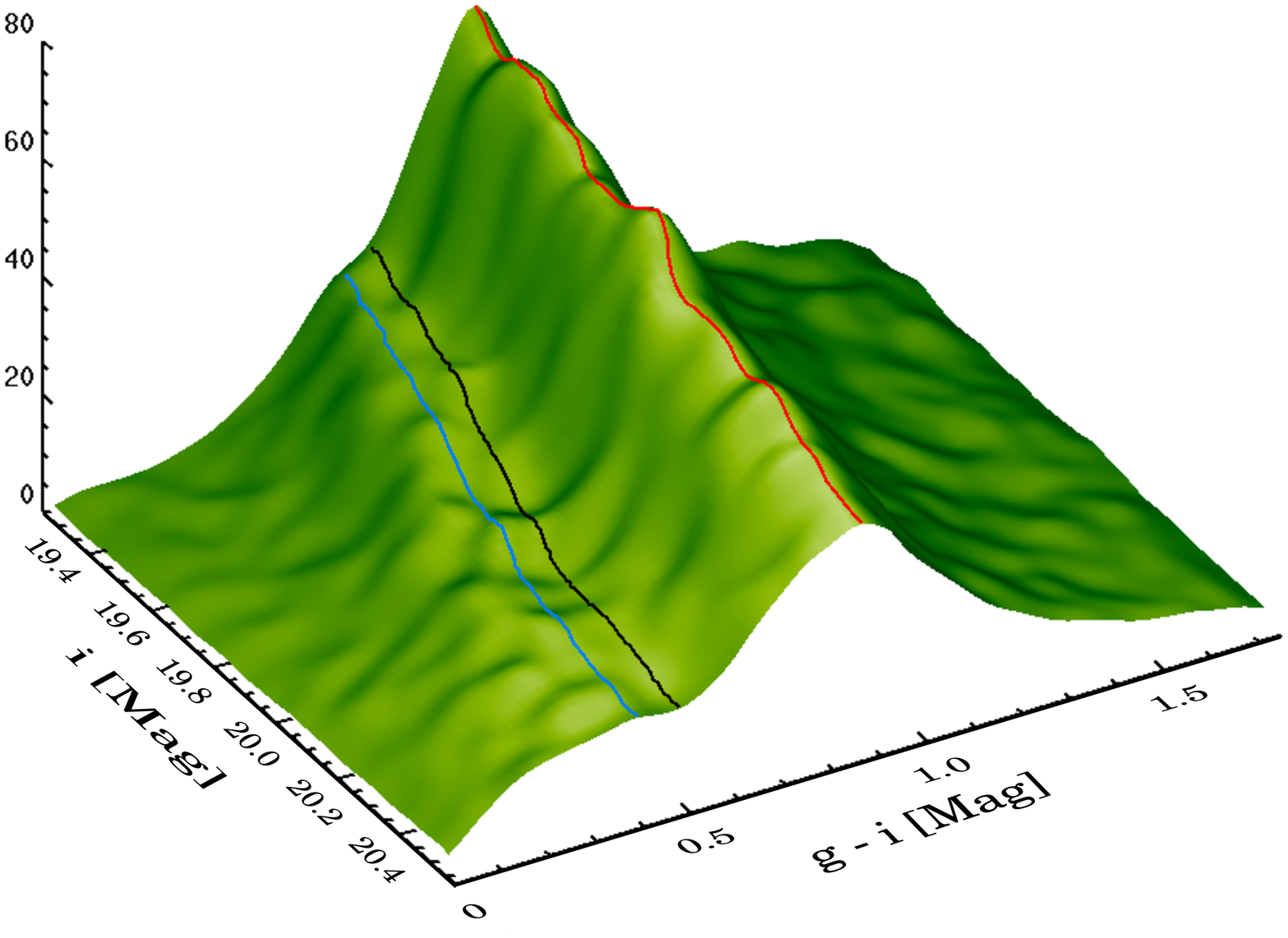} 
\caption{3D $i,\ g-i$ CMD of \omc main-sequence in the magnitude interval 19.25 $< i<$ 20.5.
The vertical axis indicates the luminosity function as described in \S7.1. The blue and red solid lines 
indicate the blue and the red main-sequence ridge lines, respectively. The black solid lines marks the valley, i.e. 
the relative minimum between the two relative maxima.}
\end{center}
\end{figure}

\begin{figure*}
\begin{center}
\label{fig10}
\includegraphics[height=0.4\textheight,width=1.0\textwidth]{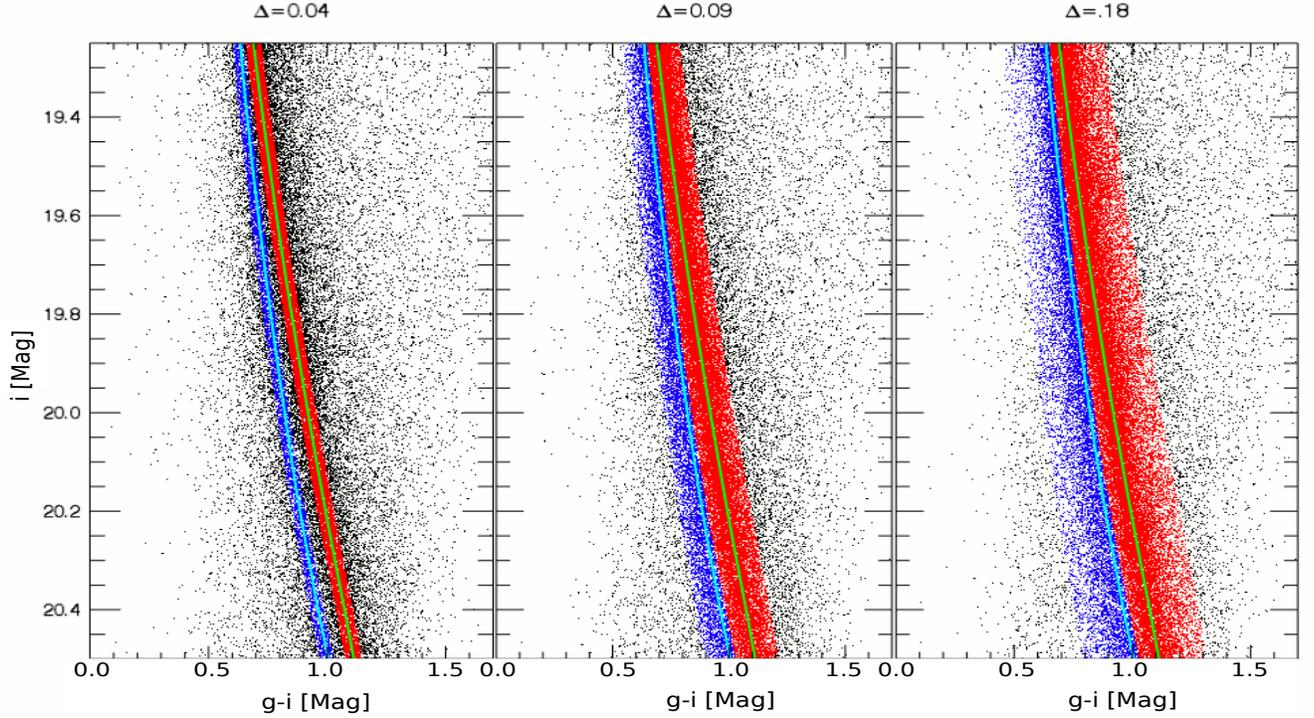} 
\caption{$i,\ g-i$ CMDs of \omc main-sequence in the magnitude interval 19.25 $< i<$ 20.5.
The three panels show blue (blue dots) and red (red) main-sequence stars selected by using
different $g-i$ color limits, $\Delta$, indicated at the top of each panel.
The cyan and green solid lines mark the blue and red main-sequence
ridge lines, respectively. See the detailed explanation in \S7.1.}
\end{center}
\end{figure*}

\begin{figure*}
\begin{center}
\label{fig11}
\includegraphics[height=0.35\textheight,width=1.0\textwidth]{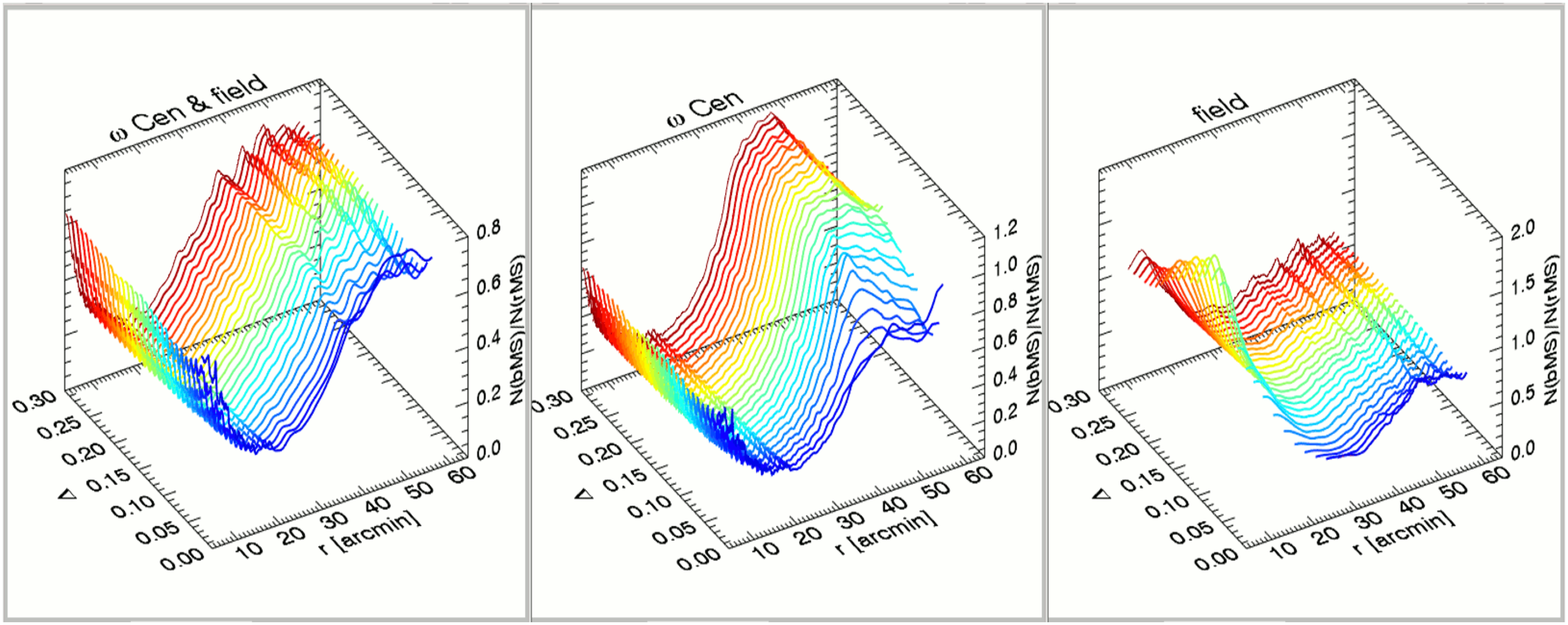} 
\caption{Ratio of bMS and rMS stars as a function of distance from the cluster center
and $g-i$ color bins used to select the stars, $\Delta$.
The three panels show the same ratio for all the stars (left), for only candidate 
cluster members (middle) and for only candidate field members (right) 
in the magnitude interval 19.25 $< i<$ 20.5.}
\end{center}
\end{figure*}

\begin{figure*}
\begin{center}
\label{fig12}
\includegraphics[height=0.35\textheight,width=0.8\textwidth]{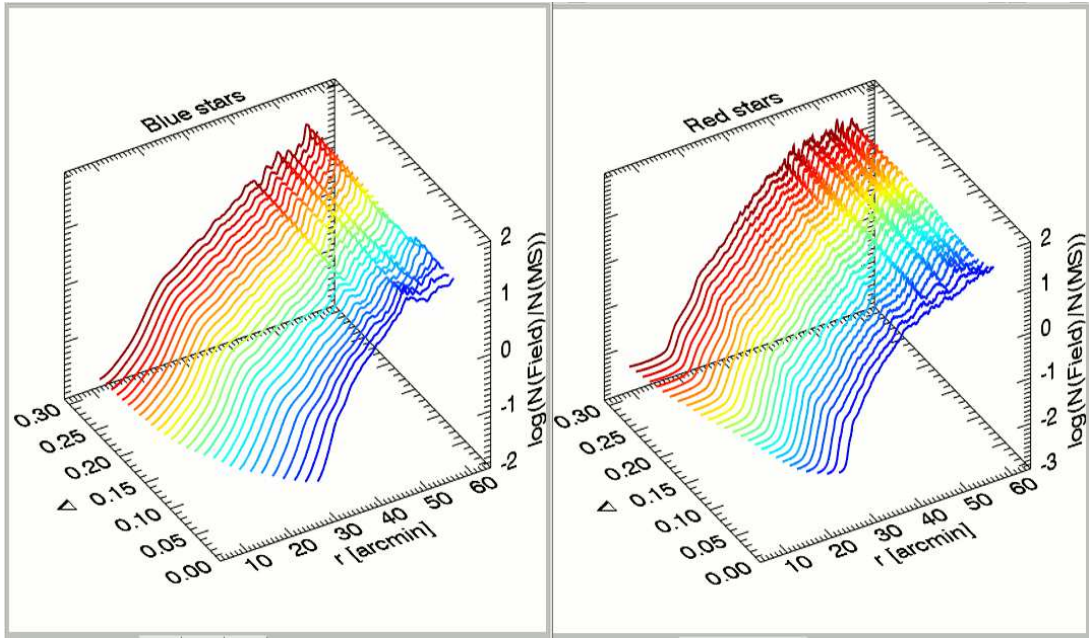} 
\caption{Logarithm of the ratio of candidate field and cluster blue (left panel) and
red (right) main-sequence stars in the magnitude interval 19.25 $< i<$ 20.5 for different 
$g-i$ color bins used to select the stars, $\Delta$.}
\end{center}
\end{figure*}

\begin{figure*}
\begin{center}
\label{fig13}
\hspace{-0.5cm}
\includegraphics[height=0.65\textheight,width=0.85\textwidth]{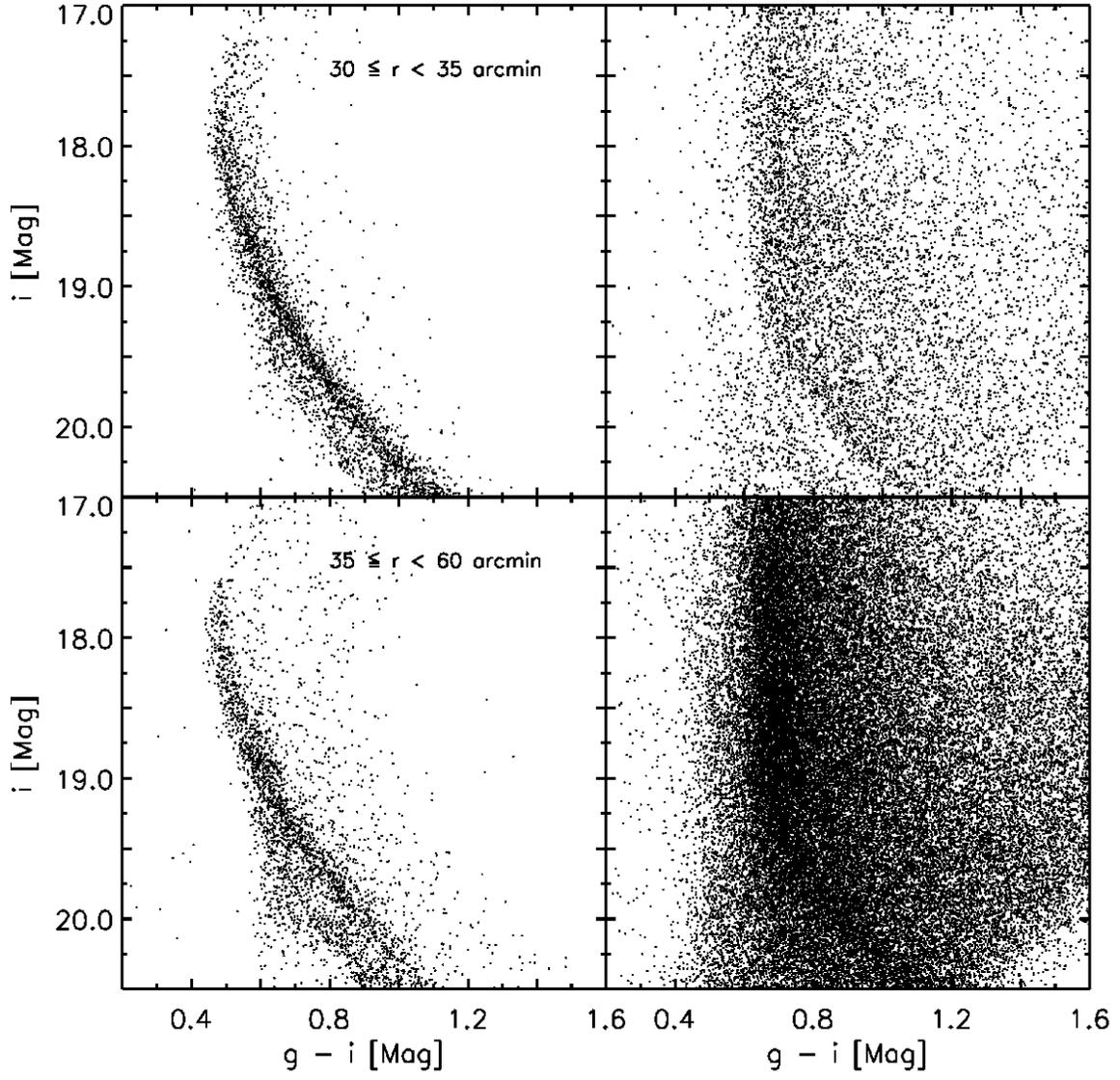} 
\caption{$i,\ g-i$ CMDs of candidate cluster (left panels) and field (right panels) stars for two 
different external radial bins.}
\end{center}
\end{figure*}

\begin{figure*}
\centering
\begin{minipage} [l]{0.48\textwidth}
\centering
\hspace{-1truecm}
\includegraphics[height=8.5cm,width=8.5cm]{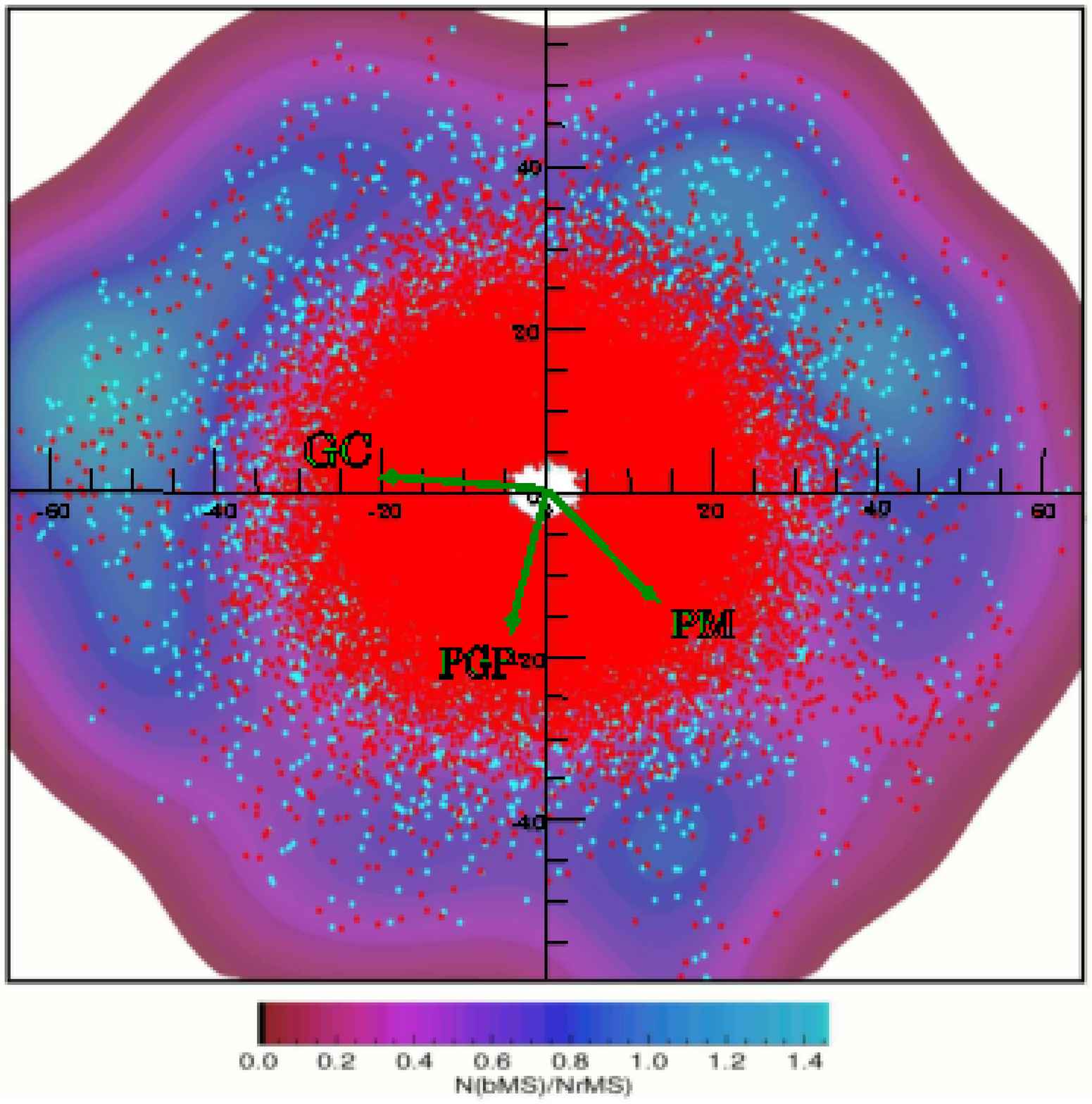}
\end{minipage}
\begin{minipage}[r]{0.48\textwidth}
\centering
\hspace{1.0truecm}
\vspace{-2.5truecm}
 \includegraphics[height=9.cm,width=9.3cm]{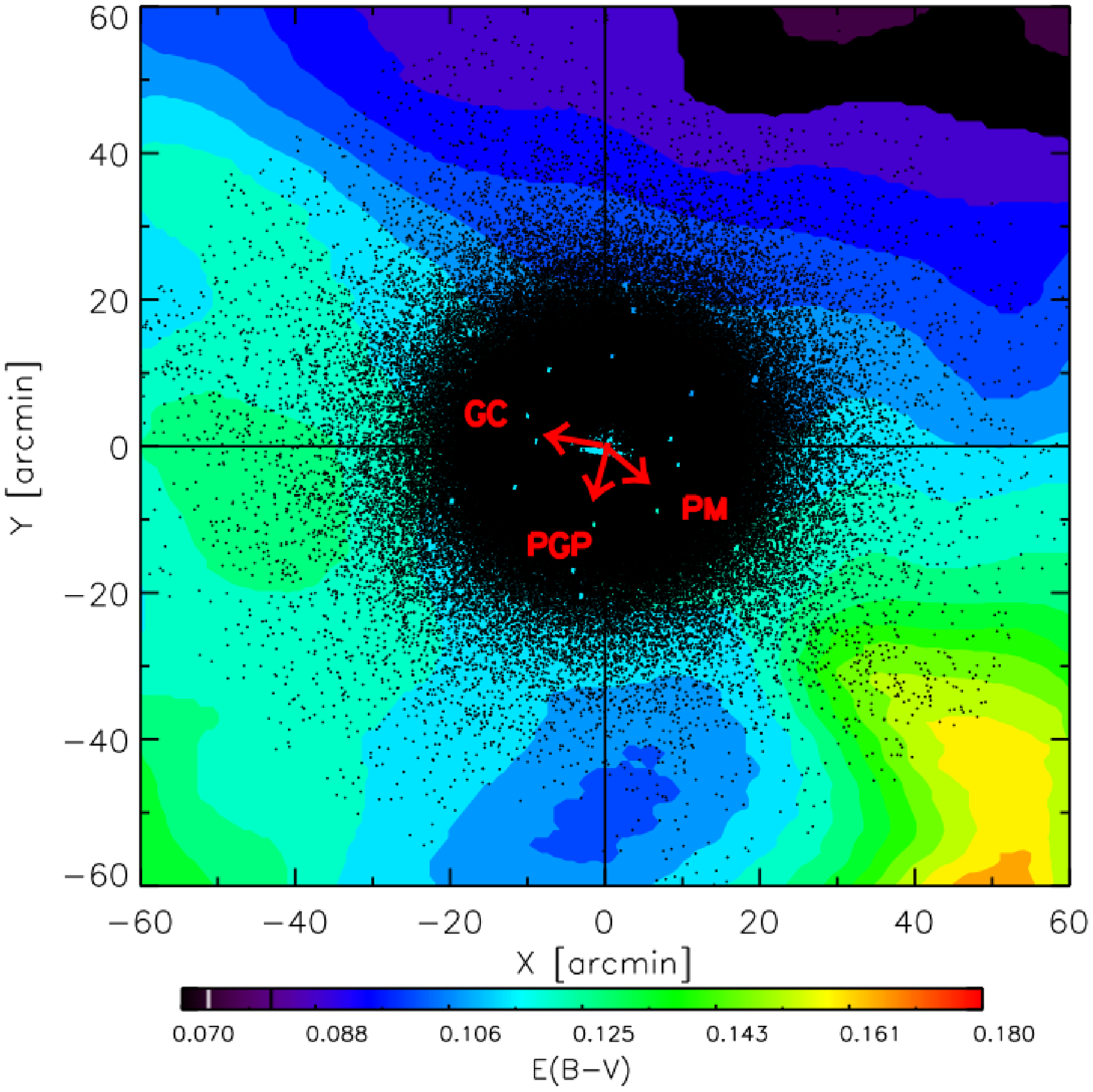}
\vspace{1.0truecm}
\end{minipage}
\caption{\label{fig14} Left - Density color map of the ratio of bMS and rMS stars as a function of star position in arcminutes.
The green arrows mark the direction of the Galactic center (GC), \omc proper motion (PM), and the direction perpendicular to the
Galactic center (PGP). The color scale is shown at the bottom.
Right - Reddening color density map for the observed region across \omc as a function of star position in arcminutes.
The direction of the GC, the cluster PM and the PGP are indicated with red arrows and the color scale is shown at the bottom.}
\end{figure*}

\subsection{Comparison with literature}

Our result is in agreement with the findings of \citet[hereinafter BE09]{bellini2009}, 
based on HST data, for distances 5 $\lesssim r \lesssim$ 10\arcmin. At larger distances 
and up to $r \approx$ 20\arcmin, i.e. the region of \omc sampled by Bellini et al.,
our ratio of bMS and rMS stars is significantly lower than the ratio found
by the quoted authors. At $r \approx$ 15\arcmin, for instance, Bellini et al. found a 
ratio of 0.36$\pm$0.04, while we find a ratio of 0.21$\pm$0.003, more than 
3$\sigma$ smaller. On the other hand, our findings do not agree with 
the study of \citet{sollima2007a} for distances smaller than $r \approx$ 10\arcmin,
while they agree very well at larger distances and up to  $r \approx$ 25\arcmin, i.e., 
the cluster region sampled by VLT photometry. 
BE09 claim that Sollima et al. ratio is lower compared to their findings 
due to the wider color range they used to select rMS stars, which would include 
unresolved binaries and members of the third MS, making the bMS and rMS 
population ratio smaller.
The different approaches used to select bMS and rMS stars might also be 
the origin of the difference we find between our ratios and Sollima et al. ratios
at small distances from the cluster center.

As far as the difference between our ratios and BE09 values for
distances larger than 10\arcmin, we have to take into account several 
circumstantial evidence. 
Our sample of rMS stars is contaminated by unresolved binaries, MS-a stars,
and marginally by blends in these more external cluster regions.
Photometric and spectroscopic analysis provide a binary
frequency for \omc of $\approx$ 5\% \citep{mayor1996, sollima2007c}, 
and MS-a stars, which are the counterpart of RGB-a stars, are less than 5\% of cluster stars 
\citep[CS07]{pancino2000}. These factors will cause an artificial increase in the star counts 
of the rMS, i.e. a decrease in the population ratio we are dealing with. 
By accounting for these factors, our population ratio would 
agree with the findings of BE09. However, these factors cannot explain the
global decreasing trend of the ratio of bMS and rMS stars 
observed with DECam data and not found by BE09. 
The number of binaries is indeed expected to decrease at increasing
distances from the cluster center, and the MS-a stars are supposed to be more centrally 
concentrated compared to metal-poor stars \citep{pancino2003, bellini2009}. 
The number of these objects and blends decreases at larger distances 
from the cluster center, with the net effect of an increase of the bMS 
and rMS ratio. However, DECam data are clearly showing that this population 
ratio is decreasing from $\approx$7\arcmin~ up to a distance of $\approx$ 25\arcmin, where it attains 
a broad minimum value of $\approx$ 0.17.

We are thus left with the following evidence:

a) the current population ratio agrees well with star counts provided by BE09 
in the innermost cluster regions (5 $\le r \le$ 10\arcmin) and attains a value $r(bMS/rMS)\approx$ 0.3--0.4;
 
b) the current star counts agree well with star counts provided by SO07 
in the cluster regions located between 10 $\le r \le$ 25\arcmin, and show a decreasing 
population ratio from a value $r(bMS/rMS)\approx$ 0.3--0.4 down to $\approx$ 0.15--0.2.

c) $r(bMS/rMS)$ is increasing for distances larger than $\approx$ 25\arcmin~ reaching a 
peak value of $\approx$ 1.2 at $\approx$ 48\arcmin

The above empirical evidence brings forward a very interesting implication. 
Spectroscopic measurements are available for 17 blue MS stars
and they suggest that the bMS is more metal-rich than the 
rMS \citep{piotto2005}. Our photometric analysis shows that the radial 
trend of this sub-population becomes more and more relevant when moving towards 
the outskirts of the cluster. This behavior is different from what it is
observed in some nearby dwarf galaxies, where more metal-rich stars are centrally 
concentrated when compared with the more metal-poor ones 
\citep{bono2010, fabrizio2015}. 
This radial trend is further supported by metallicity gradients based on 
spectroscopic measurements of nearby dwarf galaxies \citep{ho2015} and 
on photometric indices \citep{martinez2016}, suggesting either a relatively 
flat metallicity distribution or a steady decrease when moving from the 
innermost to the outermost galaxy regions. \omc seems to show an opposite
trend with a population of more metal-rich stars less concentrated compared to 
the more metal-poor population.
The reasons for the possible difference in the metallicity trend between 
\omc and nearby dwarf galaxies are not clear and new spectroscopic measurements 
to infer kinematical and abundance properties of a larger sample of bMS and rMS 
stars at different radial distances are required to better characterize their nature or nurture.

\begin{figure}
\begin{center}
\label{fig15}
\includegraphics[height=0.65\textheight,width=0.5\textwidth]{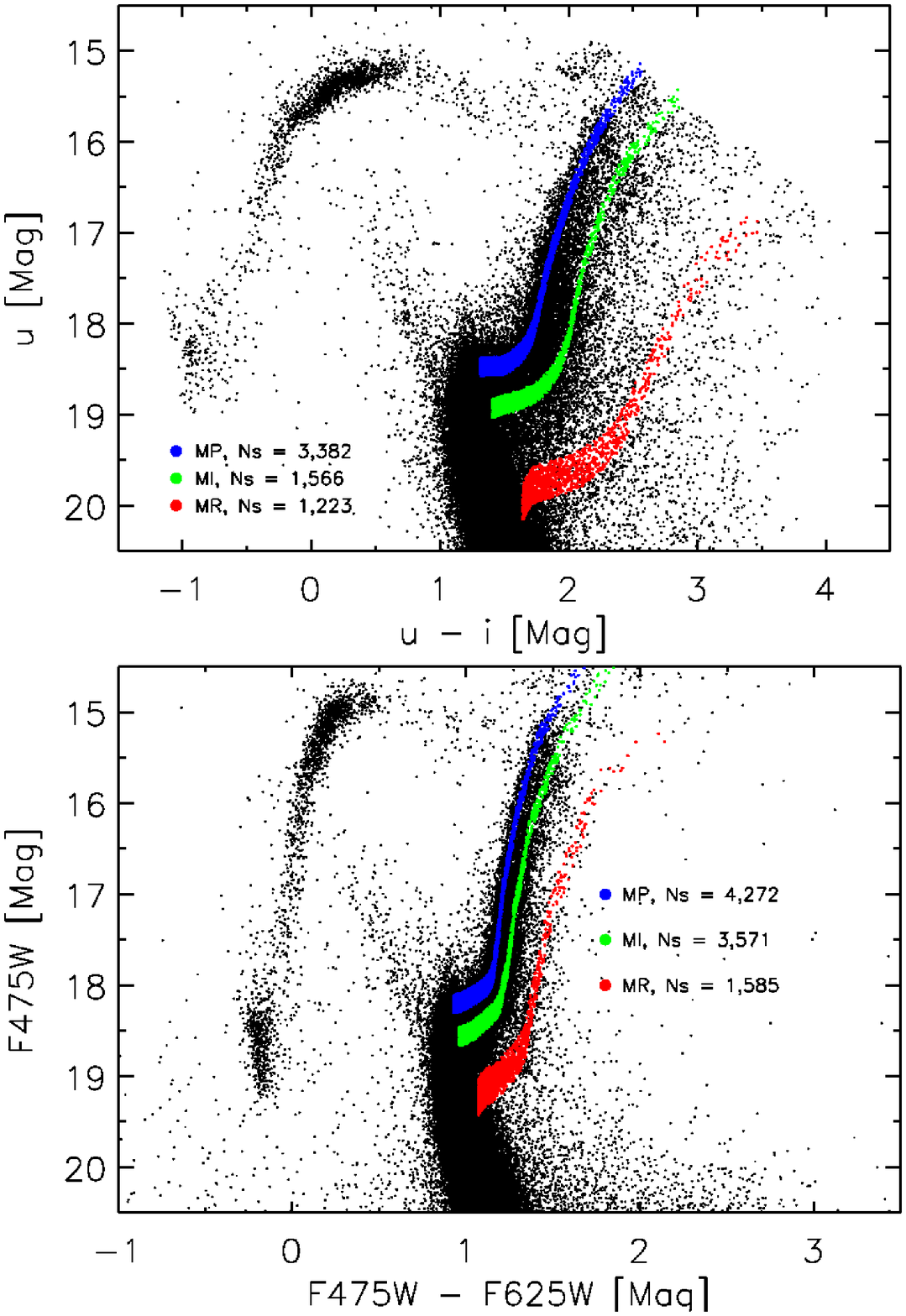} 
\caption{Top - DECam $u,\ u-i$ CMD of \omc cluster members. The selected metal-poor (MP) 
red-giant branch stars (blue dots), metal-intermediate (MI, green), and metal-rich (MR, red) are over-plotted. 
Bottom:  ACS $F475W,\ F475W - F625W$ CMD of \omc cluster members. Selected MP, MI and 
MR stars are marked.}
\end{center}
\end{figure}

\begin{figure}
\begin{center}
\label{fig16}
\includegraphics[height=0.4\textheight,width=0.5\textwidth]{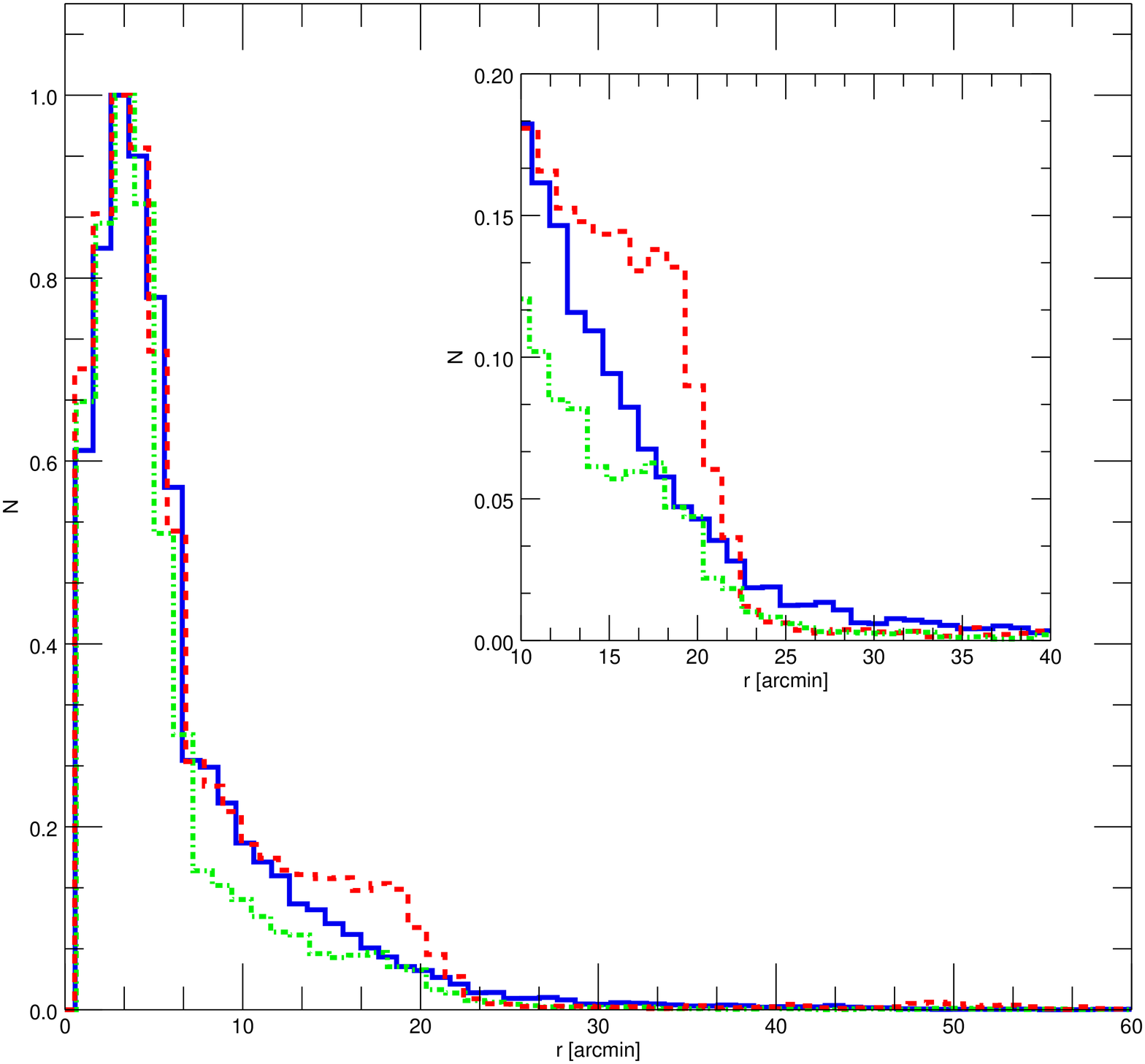} 
\vspace{-0.2cm}
\caption{Histograms of the spatial distribution of the metal-poor (blue solid line), 
metal-intermediate (green dotted-dashed), and metal-rich (red dashed) sub- and red-giant branch stars.
The inset shows a zoom of the region included in 10 $< r <$ 40\arcmin.}
\end{center}
\end{figure}

\section{The spatial distribution of red-giant branch stars}
The bMS has its counterpart in one of the multiple RGBs of \omcen, possibly at a metallicity 
intermediate for the cluster range (for more details on the correspondence between the triple MS and
the multiple sub- and red-giant branches see \citealt{bellini2010}). 
Unfortunately, it is not possible to clearly separate the different intermediate RGBs without using 
the information on the star chemical composition. For the more central regions of the cluster, 
up to $\approx$ 25\arcmin, low- and high-resolution spectroscopy and photometric metallicities 
are available. For the outskirts of the cluster, no metallicity information is available so far. 
Therefore, we decided to investigate the spatial distribution of the bluest and brightest 
SGB/RGB, the most MP sub-population in \omc according to spectroscopy,
corresponding to the rMS. We compared the properties of this sub-population to the ones of the 
faintest and reddest SGB/RGB, the most MR cluster sub-population according to 
spectroscopic measurements and corresponding to the reddest MS. 
The reddest MS, or MS-a, is difficult to separate from the rMS because it overlaps 
with its sequence of unresolved binaries. However, MS-a 
has its continuation on \omc reddest RGB, 
the $\omega$3 branch, which constitutes the most MR sub-population of the cluster based on 
spectroscopic data \citep{pancino2000, pancino2007}.
The $\omega$3 branch is well-separated from the other RGBs in the $i,\ u-i$ or $u,\ u-i$ CMDs 
and the $F475W,\ F475W - F625W$ or $F625W,\ F475W - F625W$ CMDs, where 
the temperature sensitivity is larger (see Fig.~6). Therefore, we used 
the $u,\ u-i$ CMD, and for the internal regions of the cluster, $r \lesssim$ 5\arcmin,
the $F475W,\ F475W - F625W$ CMD to select a sample of stars along the $\omega 3$ branch.
We also decided to compare the properties of $\omega 3$ stars
and of the most MP sub-population with a sample 
of stars representative of a cluster metal-intermediate (MI) sub-population.
To select candidate MP and MI stars we used the following method.
We draw two ridge lines following the MP and one MI RGB on the
$u,\ u-i$ and $F475W,\ F475W - F625W$ CMD, respectively, and 
selected stars 0.1 mag fainter and brighter than these ridge lines.
Note that the aim of this analysis is to select a sample of MP
stars and of stars with a metallicity intermediate between 
the MP and the $\omega$3 sub-population.

Fig.~15 shows DECam $u,\ u-i$ (top panel) and 
ACS $F475W,\ F475W - F625W$ CMD (bottom) with the selected sample of MP (blue dots),
MI (green) and MR (red) stars. 
The three samples of cluster stars have a similar completeness since both DECam and ACS photometric catalogs
are complete down to the turn-off level.
The total sample of MP stars includes $\approx$ 8,200 objects, while the MI and the MR one include 
$\approx$ 5,400 and $\approx$ 2,900, respectively.
Note that we are interested in investigating the spatial distribution
of these sub-populations and not in determining their absolute star counts and ratios.

Fig.~16 shows the histograms of the radial distance in arcminutes for the 
MP (solid blue line), the MI (dashed-dotted green) and the 
MR (dashed red) star samples. 
The spatial distributions of the MP and MR samples are very 
similar until $\approx$ 12\arcmin, while the fraction of MR stars increases for 
larger distances until $\approx$ 23\arcmin. 
The spatial distribution of the MI sample is different from either the MP and 
the MR spatial distributions starting from a distance of $\approx$ 8\arcmin~ from 
the cluster center.  The frequency of MI stars is lower compared to MP and MR stars, from this 
distance until the tidal radius. 
The inset of Fig.~16 shows a zoom of the radial 
distributions from 10 to 40\arcmin. It is clear how the number of MR stars increases 
compared to the MP ones starting at $\approx$ 12\arcmin~ and then start decreasing 
again at $\approx$ 23\arcmin, while the number of MI stars is always lower in this distance range.
For distances larger than $\approx$ 40\arcmin, statistics is preventing us to fully characterize the behavior of these sub-populations. 
To verify that our method to select the sample of MI red giants did not alter 
the analysis we performed the following test. We selected MI stars by moving the RGB 
ridge line 0.1 and 0.2 mag brighter and fainter on both the CMDs. 
We then compared the spatial distribution of the new sample of MI stars with those 
of the MP and MR stars. The result is the same within uncertainties, 
with the MI stars being more centrally concentrated compared to 
the MP and the MR stars.

Fig.~17 shows the cumulative radial distributions of MP, MI and MR stars. 
This plot clearly shows that MI stars are more centrally concentrated 
compared to the MP and the MR stars, while the MP stars are more concentrated than the MR
stars for distances  10 $\lesssim r  \lesssim$ 20\arcmin. 
Moreover, the $\omega 3$ sub-population has a more extended spatial 
distribution when considering distances larger than 10\arcmin. This result confirms 
the findings of CS07, where an increase of $\omega$3 star counts compared to the 
other RGs was observed with increasing distance from the cluster center. In particular, 
the fraction of $\omega$3 stars increases from $\approx$ 3\% at $r \approx$ 8\arcmin~ 
to $\approx$ 7\% for larger distances (for more details please see their Table~3). 

The current results also agree with previous findings by \citet{hilker2000} based on \strom 
photometric metallicities for a sample of \omc RGs. These authors 
showed that more metal-rich RGs are more concentrated compared to more metal-poor stars
within a radius of 10\arcmin~ from the cluster center.  \citet{sollima2005a}, based on 
photometry of a sample of RGs for a field of view of $\approx$ 0.2$\times$0.2$\deg$ across 
the cluster, found a similar result. All these previous studies were based on photometric 
catalogs covering a radial distance until $\approx$ 20\arcmin.
DECam data allowed us for the first time to analyze the spatial 
distribution of \omc RGB sub-populations across a much larger portion of the cluster
and to disclose the peculiar behavior of $\omega 3$ branch stars.

We also investigated for the presence of a spatial segregation of the MP, MI and MR 
star sample.
Table~6 lists the ratio of MR, MI and MP stars, $N(MR)/N(MP)$ and $N(MI)/N(MP)$, 
in the four quadrants of \omcen, i.e. NW, NE, SW and SE. For the MR to MP star ratio,
values are in agreement within uncertainties in the three NW, SW and NE 
quadrants, while a deficiency of MP stars is present in the SE quadrant. 
The MI to MP star ratio show a clear West-East asymmetry, due to the decrease of
MP stars in the Southern quadrant of the cluster, while MI stars are more numerous in
the NW, NE, and SE quadrants of \omcen.
The same table lists the number of MP, MI and MR stars in the four different regions.
A clear asymmetry is present, with the MP being much more numerous in the two Northern quadrants, 
while MR and MI stars have a deficiency in the SW quadrant of \omcen.

\citet{jurcsik1998} showed that more metal-rich stars in \omc ($[Fe/H] \ge -1.25$) are segregated in the Southern part of the cluster,
while the metal-poor ($[Fe/H] \le -1.75$) stars in the Northern. The centroid of these two groups are $\approx$ 6\arcmin~ apart.
The segregation of MR stars in the Southern half of \omc was confirmed by \citet{hilker2000}, while they did 
not find an equivalent segregation for the MP sub-population. 
Our data seem to support a different distribution of MP, MI and MR stars in \omcen, 
and a clear excess of MP stars in the Northern half of the cluster and a deficiency of MI and MR stars in the SW quadrant.
However, we do not have homogenous abundance measurements for MP, MI and MR stars up to \omc tidal radius.
Spectroscopic measurements for stars belonging to the different sub-populations in the
outskirts of the cluster are now needed to better characterize the spatial distribution of MP, MI and MR RGB stars in \omcen.


\begin{table*}
\caption{Star counts and population ratio for bMS and rMS candidate \omc and field stars.}\label{table:5}     
\begin{tabular}{ c c c c c c c}       
\hline\hline                
Distance & $N(rMS)$ & $N(NbMS)$ & $N(rMS)_{field}$  & $N(bMS)_{field}$  & $r(bMS/rMS)$  & $r(bMs/rMs)_{field}$  \\    
(arcmin)  &                  &                     &                              &                              &                          &                                 \\    
\hline                      
\hline
 7.5   &    21128     &     7672        &               69          &               50       &        0.36$\pm$0.003    &         0.72$\pm$0.13     \\
12.5   &   26778     &     6921        &               23          &               93       &        0.25$\pm$0.003    &         4.04$\pm$0.15     \\    
17.5   &    16271    &     3053        &               54          &               82       &        0.19$\pm$0.004    &         1.52$\pm$0.03     \\    
22.5   &      8051    &     1340        &             314          &             200       &        0.17$\pm$0.005    &         0.64$\pm$0.02     \\
27.5   &      3360    &       717        &             663          &             270       &        0.21$\pm$0.009    &         0.41$\pm$0.01     \\    
32.5   &      1308    &       512        &            1171          &            384       &        0.39$\pm$0.02      &         0.33$\pm$0.01     \\    
37.5   &        470    &       334        &            1349          &            588       &        0.71$\pm$0.05      &         0.44$\pm$0.01     \\    
42.5   &        199    &       193        &            1401          &            708       &        0.97$\pm$0.10      &         0.50$\pm$0.01     \\    
47.5   &        115    &       143        &            1387          &            873       &        1.24$\pm$0.15       &        0.63$\pm$0.01     \\    
52.5   &          93    &         93        &            1383          &            915       &        1.00$\pm$0.15       &        0.66$\pm$0.01     \\    
57.5   &          78    &         59        &            1166          &            778       &        0.76$\pm$0.13       &        0.67$\pm$0.02     \\    
\hline
\hline
\end{tabular}
\end{table*}



\begin{table}
\caption{Number of metal-poor (MP) and metal-rich (MR) sub- and red-giant branch stars in the
different quadrants of \omcen.}\label{table:6}     
\begin{tabular}{ l c c c c  c}       
\hline\hline                
 & $N(MR)/N(MP)$ & $N(MI)/N(MP)$ & $N(MP)$  & $N(MI)$ & $N(MR)$  \\    
\hline                      
\hline
NW   & 0.34$\pm$0.02      & 0.57$\pm$0.02   & 2651$\pm$51  & 1504$\pm$39 & 829$\pm$29 \\
\hline                      
SW  &  0.35$\pm$0.02      &  0.57$\pm$0.02  & 1837$\pm$43  & 1056$\pm$32 & 636$\pm$25 \\
\hline
NE  &  0.38$\pm$0.02       &  0.72$\pm$0.02 &  1915$\pm$44  & 1377$\pm$37 & 761$\pm$28 \\
\hline                      
SE  &  0.41$\pm$0.02       &  0.81$\pm$0.03 & 1616$\pm$40  & 1318$\pm$36  & 682$\pm$26 \\
\hline
\hline
\end{tabular}
\end{table}


\section{Summary and conclusions}\label{concl}

We presented multi-band photometry of \omc for a total FoV of $\approx$ 2$\times$2$\deg$ 
across the cluster. 
Images were collected with the wide field camera DECam and combined with ACS 
data for the crowded regions of the cluster core. 
The availability of the $u$-band photometry allowed us to use a new method based
on color-color-magnitude diagrams to separate cluster and field stars. 
We ended up with a final photometric catalog of $\approx$ 1.7 million cluster members, 
including photometry in seven filters, namely $F475W, F625W, F658N, u, g, r, i$.
To our knowledge, this is the largest multi-band data set ever collected for a Galactic 
globular cluster and covering the widest FoV after our ACS-WFI catalog published in CS07.

DECam precise photometry allowed us to observe the split along \omc MS and 
to show that it is present at all distances from the cluster center.
The bMS is well-separated from the rMS in the magnitude 
range 19.0 $\le i \le$ 21.0.  The $g - i$ color distance between the two sequences is changing with 
magnitude and reaches a maximum of -0.18 mag at $i \approx$ 21 mag. 
The color separation between the rMS and the bMS is the same, within the uncertainties, 
at all distances from \omc center, ranging from $\approx$ 0.07 to $\approx$ 0.18 mag according to the magnitude interval.

DECam data allowed us for the first time to analyze the spatial 
distribution of the different \omc sub-populations across the cluster until the nominal 
tidal radius of 57\arcmin. In particular, we were able to investigate the spatial distribution of the two cluster 
main sequences. We found that stars belonging to the bMS are more centrally concentrated compared to stars 
belonging to the rMS up to a distance of $\approx$ 25\arcmin.  The frequency of bMS stars is then 
steadily increasing up to $\approx$ 50\arcmin, with the ratio of bMS and rMS stars being larger than 1.
The ratio of bMS to rMS stars shows an asymmetric clumpy distribution across the cluster, with 
an excess of bMS stars in the Northern half. 
The over-density of bMS stars in the cluster North-East quadrant is pointing towards the Galactic center, 
suggesting a connection between the spatial distribution of bMS stars and 
\omc dynamical evolution.

Unfortunately, our photometry does not allow us to identify the continuation of the bMS on the sub- and
red-giant branch phases. We then analyzed the spatial distribution of a sample of MP
and MI stars selected along the sub- and red-giant branch by using ridge lines 
in the $u,\ u-i$ and $F475W,\ F475W - F625W$ CMDs. Moreover, stars belonging 
to the $\omega$3 branch, the most MR sub-population in the cluster according to spectroscopy,
were also selected. The three samples show a different spatial distribution; MI stars are more 
concentrated compared to MP and MR ones, while MP star are more concentrated than 
MR stars for distances in the interval 10 $\lesssim r  \lesssim$ 20\arcmin. 
Data clearly show that the $\omega 3$ sub-population has a more extended spatial 
distribution when considering distances larger than 10\arcmin. This result confirms the findings of CS07, where 
an increase of the number of $\omega$3 stars compared to the other RGs
with increasing distance from the cluster center was found. 
Star counts of the MP, MI and MR samples show that a deficiency of MI and MR stars 
the SW quadrant of \omc is present. Moreover, MP stars are more numerous 
in the Northern half of the cluster. 

Stellar populations with different metallicities and age show different spatial distributions 
with the more metal-rich sub-populations being more centrally concentrated in some nearby 
dwarf spheroidal galaxies such as Carina, Sculptor, Fornax \citep{monelli2003, delpino2013, fabrizio2015, ho2015, martinez2016}.
The same behavior is observed in a Galactic globular cluster presenting a significant spread in iron abundance, 
Terzan~5. The metal-rich sub-population in this cluster is more centrally concentrated compared to the metal-poor one \citep{ferraro2016}.
The case of \omc seems to be different from both known Local Group dwarf spheroidal galaxies and 
clusters presenting stellar populations with different metallicities. \omc MI selected RGs are indeed more 
centrally concentrated compared to the MP ones, but the $\omega 3$ sub-population shows 
a more extended spatial distribution. The bMS stars, spectroscopically claimed to be more 
metal-rich compared to rMs stars, also show a more extended and very asymmetric spatial distribution.
A few main conclusions can then be drawn:

\begin{itemize}

\item \omc hosts a metal-intermediate RGB sub-population that behaves as the 
metal-intermediate and metal-rich stellar populations in some nearby dwarf 
spheroidal galaxies and Terzan~5, being more centrally concentrated compared to the cluster 
metal-poor RGB sub-population. This stellar sub-population is peculiar compared to 
the typical second generation of stars present in other globular clusters, 
presenting not only a different iron abundance and its own chemical anti-correlations, 
according to spectroscopic analyses \citep{gratton2011, marino2011, marino2012}, 
but also a very different spatial distribution;

\item \omc bMS stars are more centrally concentrated compared to rMS stars 
up to a distance of $\approx$ 25\arcmin. The frequency of bMS stars, supposedly more 
metal-rich than the rMS stars according to spectroscopic measurements, 
steadily increases at larger distances, outnumbering the 
rMS stars until approximately the cluster tidal radius. 
Their spatial distribution is asymmetric and clumpy, with an excess of bMS stars
in the direction of the Galactic center;

\item \omc hosts a metal-rich sub-population making up the cluster third MS, MS-a, 
and the reddest and faintest RGB, the $\omega$3 branch. These stars are more centrally 
concentrated compared to more metal-poor stars up to a distance of 
$\approx$ 10\arcmin~ from the cluster center, and their frequency increases at larger distances, 
showing a more extended spatial distribution;

\item The behavior of the bMS and the $\omega$3 branch is similar, being both
sub-populations more concentrated in the central regions and more extended in the 
outskirts of the cluster. This results suggests a possible
common origin for these stellar sub-populations.

\end{itemize}

The current photometric catalog, thanks to its accuracy and spatial coverage, allowed 
us to disclose the complex spatial distribution of different stellar sub-populations in \omcen.
These results, if confirmed, will make \omc the only stellar system known to have more metal-rich
stars with a more extended spatial distribution compared to more metal-poor stars.

Further data are now needed to solve the \omc puzzle.
The current photometry combined with abundance and radial velocity measurements for stars
of the different stellar sub-populations across the entire cluster will allow us to 
better understand the origin of \omcen.

Further homogeneous photometric data are also needed to better characterize the behavior of 
the different stellar sub-populations until and beyond the nominal tidal radius of 57\arcmin.
We have an approved DECam proposal to observe an area around \omc beyond its nominal 
tidal radius. The new data will allow us to better characterize the spatial distribution 
of \omc different stellar sub-populations.
In particular, we are interested in confirming the more extended 
spatial distribution of the bMS and the $\omega$3 branch stars.
With homogenous photometry covering a FoV of at least 3$\times$3\deg~ across the cluster and including the $u$-filter, we will be also able to
confirm the presence of stellar over-densities tracing tidal tails around \omcen, previously found by \citet{marconi2014}, 
and to detect new stellar debris if present.

\begin{figure}
\begin{center}
\label{fig17}
\includegraphics[height=0.4\textheight,width=0.5\textwidth]{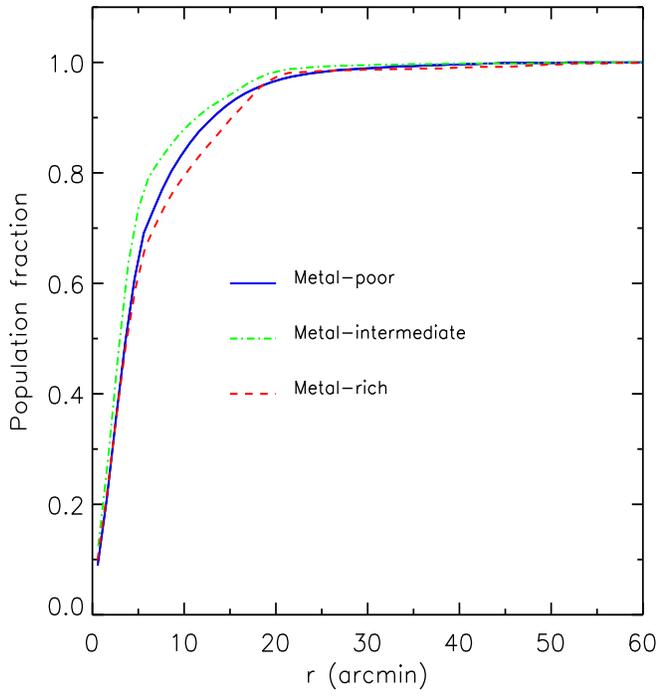} 
\caption{Cumulative radial distributions of metal-poor (blue solid line), metal-intermediate (green dashed-dotted), 
and metal-rich (red dashed) sub- and red-giant branch stars.}
\end{center}
\end{figure}

\acknowledgments
We thank the anonymous referee for very helpful suggestions which led 
to an improved version of the paper.



\clearpage
\bibliographystyle{aa}

\bibliography{calamida}

\begin{thebibliography}{57}
\expandafter\ifx\csname natexlab\endcsname\relax\def\natexlab#1{#1}\fi

\bibitem[{{Anderson}(2002)}]{anderson2002}
{Anderson}, J. 2002, in Astronomical Society of the Pacific Conference Series,
  Vol. 265, Omega Centauri, A Unique Window into Astrophysics, ed. F.~{van
  Leeuwen}, J.~D. {Hughes}, \& G.~{Piotto}, 87

\bibitem[{{Bedin} {et~al.}(2004){Bedin}, {Piotto}, {Anderson}, {Cassisi},
  {King}, {Momany}, \& {Carraro}}]{bedin2004}
{Bedin}, L.~R., {Piotto}, G., {Anderson}, J., {et~al.} 2004, \apjl, 605, L125

\bibitem[{{Bellini} {et~al.}(2010){Bellini}, {Bedin}, {Piotto}, {Milone},
  {Marino}, \& {Villanova}}]{bellini2010}
{Bellini}, A., {Bedin}, L.~R., {Piotto}, G., {et~al.} 2010, \aj, 140, 631

\bibitem[{{Bellini} {et~al.}(2009){Bellini}, {Piotto}, {Bedin}, {King},
  {Anderson}, {Milone}, \& {Momany}}]{bellini2009}
{Bellini}, A., {Piotto}, G., {Bedin}, L.~R., {et~al.} 2009, \aap, 507, 1393

\bibitem[{{Bono} {et~al.}(2010){Bono}, {Stetson}, {Walker}, {Monelli},
  {Fabrizio}, {Pietrinferni}, {Brocato}, {Buonanno}, {Caputo}, {Cassisi},
  {Castellani}, {Cignoni}, {Corsi}, {Dall'Ora}, {Degl'Innocenti}, {Fran{\c
  c}ois}, {Ferraro}, {Iannicola}, {Nonino}, {Prada Moroni}, {Pulone}, {Smith},
  \& {Thevenin}}]{bono2010}
{Bono}, G., {Stetson}, P.~B., {Walker}, A.~R., {et~al.} 2010, \pasp, 122, 651

\bibitem[{{Braga} {et~al.}(2016){Braga}, {Stetson}, {Bono}, {Dall'Ora},
  {Ferraro}, {Fiorentino}, {Freyhammer}, {Iannicola}, {Marengo}, {Neeley},
  {Valenti}, {Buonanno}, {Calamida}, {Castellani}, {da Silva},
  {Degl'Innocenti}, {Di Cecco}, {Fabrizio}, {Freedman}, {Giuffrida}, {Lub},
  {Madore}, {Marconi}, {Marinoni}, {Matsunaga}, {Monelli}, {Persson},
  {Piersimoni}, {Pietrinferni}, {Prada-Moroni}, {Pulone}, {Stellingwerf},
  {Tognelli}, \& {Walker}}]{braga2016}
{Braga}, V.~F., {Stetson}, P.~B., {Bono}, G., {et~al.} 2016, \aj, 152, 170

\bibitem[{{Calamida} {et~al.}(2009){Calamida}, {Bono}, {Stetson}, {Freyhammer},
  {Piersimoni}, {Buonanno}, {Caputo}, {Cassisi}, {Castellani}, {Corsi},
  {Dall'Ora}, {Degl'Innocenti}, {Ferraro}, {Grundahl}, {Hilker}, {Iannicola},
  {Monelli}, {Nonino}, {Patat}, {Pietrinferni}, {Prada Moroni}, {Primas},
  {Pulone}, {Richtler}, {Romaniello}, {Storm}, \& {Walker}}]{calamida2009}
{Calamida}, A., {Bono}, G., {Stetson}, P.~B., {et~al.} 2009, \apj, 706, 1277

\bibitem[{{Calamida} {et~al.}(2005){Calamida}, {Stetson}, {Bono}, {Freyhammer},
  {Grundahl}, {Hilker}, {Andersen}, {Buonanno}, {Cassisi}, {Corsi}, {Dall'Ora},
  {Del Principe}, {Ferraro}, {Monelli}, {Munteanu}, {Nonino}, {Piersimoni},
  {Pietrinferni}, {Pulone}, \& {Richtler}}]{calamida2005}
{Calamida}, A., {Stetson}, P.~B., {Bono}, G., {et~al.} 2005, \apjl, 634, L69

\bibitem[{{Cannon} \& {Stobie}(1973)}]{cannonstobie1973}
{Cannon}, R.~D. \& {Stobie}, R.~S. 1973, \mnras, 162, 207

\bibitem[{{Cardelli} {et~al.}(1989){Cardelli}, {Clayton}, \&
  {Mathis}}]{cardelli89}
{Cardelli}, J.~A., {Clayton}, G.~C., \& {Mathis}, J.~S. 1989, \apj, 345, 245

\bibitem[{{Castellani} {et~al.}(2007){Castellani}, {Calamida}, {Bono},
  {Stetson}, {Freyhammer}, {Degl'Innocenti}, {Moroni}, {Monelli}, {Corsi},
  {Nonino}, {Buonanno}, {Caputo}, {Castellani}, {Dall'Ora}, {Del Principe},
  {Ferraro}, {Iannicola}, {Piersimoni}, {Pulone}, \& {Vuerli}}]{castellani2007}
{Castellani}, V., {Calamida}, A., {Bono}, G., {et~al.} 2007, \apj, 663, 1021

\bibitem[{{Cutri} {et~al.}(2003){Cutri}, {Skrutskie}, {van Dyk}, {Beichman},
  {Carpenter}, {Chester}, {Cambresy}, {Evans}, {Fowler}, {Gizis}, {Howard},
  {Huchra}, {Jarrett}, {Kopan}, {Kirkpatrick}, {Light}, {Marsh}, {McCallon},
  {Schneider}, {Stiening}, {Sykes}, {Weinberg}, {Wheaton}, {Wheelock}, \&
  {Zacarias}}]{cutri2003}
{Cutri}, R.~M., {Skrutskie}, M.~F., {van Dyk}, S., {et~al.} 2003, VizieR Online
  Data Catalog, 2246

\bibitem[{{Dauphole} {et~al.}(1996){Dauphole}, {Geffert}, {Colin}, {Ducourant},
  {Odenkirchen}, \& {Tucholke}}]{dauphole1996}
{Dauphole}, B., {Geffert}, M., {Colin}, J., {et~al.} 1996, \aap, 313, 119

\bibitem[{{del Pino} {et~al.}(2013){del Pino}, {Hidalgo}, {Aparicio},
  {Gallart}, {Carrera}, {Monelli}, {Buonanno}, \& {Marconi}}]{delpino2013}
{del Pino}, A., {Hidalgo}, S.~L., {Aparicio}, A., {et~al.} 2013, \mnras, 433,
  1505

\bibitem[{{Di Cecco} {et~al.}(2015){Di Cecco}, {Bono}, {Prada Moroni},
  {Tognelli}, {Allard}, {Stetson}, {Buonanno}, {Ferraro}, {Iannicola},
  {Monelli}, {Nonino}, \& {Pulone}}]{dicecco2015}
{Di Cecco}, A., {Bono}, G., {Prada Moroni}, P.~G., {et~al.} 2015, \aj, 150, 51

\bibitem[{{Fabrizio} {et~al.}(2015){Fabrizio}, {Nonino}, {Bono}, {Primas},
  {Th{\'e}venin}, {Stetson}, {Cassisi}, {Buonanno}, {Coppola}, {da Silva},
  {Dall'Ora}, {Ferraro}, {Genovali}, {Gilmozzi}, {Iannicola}, {Marconi},
  {Monelli}, {Romaniello}, \& {Walker}}]{fabrizio2015}
{Fabrizio}, M., {Nonino}, M., {Bono}, G., {et~al.} 2015, \aap, 580, A18

\bibitem[{{Ferraro} {et~al.}(2002){Ferraro}, {Bellazzini}, \&
  {Pancino}}]{ferraro2002}
{Ferraro}, F.~R., {Bellazzini}, M., \& {Pancino}, E. 2002, \apjl, 573, L95

\bibitem[{{Ferraro} {et~al.}(2016){Ferraro}, {Massari}, {Dalessandro},
  {Lanzoni}, {Origlia}, {Rich}, \& {Mucciarelli}}]{ferraro2016}
{Ferraro}, F.~R., {Massari}, D., {Dalessandro}, E., {et~al.} 2016, \apj, 828,
  75

\bibitem[{{Ferraro} {et~al.}(2004){Ferraro}, {Sollima}, {Pancino},
  {Bellazzini}, {Straniero}, {Origlia}, \& {Cool}}]{ferraro2004}
{Ferraro}, F.~R., {Sollima}, A., {Pancino}, E., {et~al.} 2004, \apjl, 603, L81

\bibitem[{{Fukugita} {et~al.}(1996){Fukugita}, {Ichikawa}, {Gunn}, {Doi},
  {Shimasaku}, \& {Schneider}}]{fukugita1996}
{Fukugita}, M., {Ichikawa}, T., {Gunn}, J.~E., {et~al.} 1996, \aj, 111, 1748

\bibitem[{{Gratton} {et~al.}(2011){Gratton}, {Johnson}, {Lucatello}, {D'Orazi},
  \& {Pilachowski}}]{gratton2011}
{Gratton}, R.~G., {Johnson}, C.~I., {Lucatello}, S., {D'Orazi}, V., \&
  {Pilachowski}, C. 2011, \aap, 534, A72

\bibitem[{{Hilker} \& {Richtler}(2000)}]{hilker2000}
{Hilker}, M. \& {Richtler}, T. 2000, \aap, 362, 895

\bibitem[{{Ho} {et~al.}(2015){Ho}, {Geha}, {Tollerud}, {Zinn}, {Guhathakurta},
  \& {Vargas}}]{ho2015}
{Ho}, N., {Geha}, M., {Tollerud}, E.~J., {et~al.} 2015, \apj, 798, 77

\bibitem[{{Johnson} \& {Pilachowski}(2010)}]{johnson2010}
{Johnson}, C.~I. \& {Pilachowski}, C.~A. 2010, \apj, 722, 1373

\bibitem[{{Jurcsik}(1998)}]{jurcsik1998}
{Jurcsik}, J. 1998, \apjl, 506, L113

\bibitem[{{Kayser} {et~al.}(2006){Kayser}, {Hilker}, {Richtler}, \&
  {Willemsen}}]{kayser2006}
{Kayser}, A., {Hilker}, M., {Richtler}, T., \& {Willemsen}, P.~G. 2006, \aap,
  458, 777

\bibitem[{{Leon} {et~al.}(2000){Leon}, {Meylan}, \& {Combes}}]{leon2000}
{Leon}, S., {Meylan}, G., \& {Combes}, F. 2000, \aap, 359, 907

\bibitem[{{Li} {et~al.}(2016){Li}, {DePoy}, {Marshall}, {Tucker}, {Kessler},
  {Annis}, {Bernstein}, {Boada}, {Burke}, {Finley}, {James}, {Kent}, {Lin},
  {Marriner}, {Mondrik}, {Nagasawa}, {Rykoff}, {Scolnic}, {Walker}, {Wester},
  {Abbott}, {Allam}, {Benoit-L{\'e}vy}, {Bertin}, {Brooks}, {Capozzi}, {Carnero
  Rosell}, {Carrasco Kind}, {Carretero}, {Crocce}, {Cunha}, {D'Andrea}, {da
  Costa}, {Desai}, {Diehl}, {Doel}, {Flaugher}, {Fosalba}, {Frieman},
  {Gaztanaga}, {Goldstein}, {Gruen}, {Gruendl}, {Gutierrez}, {Honscheid},
  {Kuehn}, {Kuropatkin}, {Maia}, {Melchior}, {Miller}, {Miquel}, {Mohr},
  {Neilsen}, {Nichol}, {Nord}, {Ogando}, {Plazas}, {Romer}, {Roodman}, {Sako},
  {Sanchez}, {Scarpine}, {Schubnell}, {Sevilla-Noarbe}, {Smith},
  {Soares-Santos}, {Sobreira}, {Suchyta}, {Tarle}, {Thomas}, {Vikram}, \& {DES
  Collaboration}}]{li2016}
{Li}, T.~S., {DePoy}, D.~L., {Marshall}, J.~L., {et~al.} 2016, \aj, 151, 157

\bibitem[{{Marconi} {et~al.}(2014){Marconi}, {Musella}, {Di Criscienzo},
  {Cignoni}, {Dall'Ora}, {Bono}, {Ripepi}, {Brocato}, {Raimondo}, {Grado},
  {Limatola}, {Coppola}, {Moretti}, {Stetson}, {Calamida}, {Cantiello},
  {Capaccioli}, {Cappellaro}, {Cioni}, {Degl'Innocenti}, {De Martino}, {Di
  Cecco}, {Ferraro}, {Iannicola}, {Prada Moroni}, {Silvotti}, {Buonanno},
  {Getman}, {Napolitano}, {Pulone}, \& {Schipani}}]{marconi2014}
{Marconi}, M., {Musella}, I., {Di Criscienzo}, M., {et~al.} 2014, \mnras, 444,
  3809

\bibitem[{{Marino} {et~al.}(2012){Marino}, {Milone}, {Piotto}, {Cassisi},
  {D'Antona}, {Anderson}, {Aparicio}, {Bedin}, {Renzini}, \&
  {Villanova}}]{marino2012}
{Marino}, A.~F., {Milone}, A.~P., {Piotto}, G., {et~al.} 2012, \apj, 746, 14

\bibitem[{{Marino} {et~al.}(2011){Marino}, {Milone}, {Piotto}, {Villanova},
  {Gratton}, {D'Antona}, {Anderson}, {Bedin}, {Bellini}, {Cassisi}, {Geisler},
  {Renzini}, \& {Zoccali}}]{marino2011}
{Marino}, A.~F., {Milone}, A.~P., {Piotto}, G., {et~al.} 2011, \apj, 731, 64

\bibitem[{{Mart{\'{\i}}nez-V{\'a}zquez}
  {et~al.}(2016){Mart{\'{\i}}nez-V{\'a}zquez}, {Monelli}, {Gallart}, {Bono},
  {Bernard}, {Stetson}, {Ferraro}, {Walker}, {Dall'Ora}, {Fiorentino}, \&
  {Iannicola}}]{martinez2016}
{Mart{\'{\i}}nez-V{\'a}zquez}, C.~E., {Monelli}, M., {Gallart}, C., {et~al.}
  2016, \mnras, 461, L41

\bibitem[{{Mayor} {et~al.}(1996){Mayor}, {Duquennoy}, {Udry}, {Andersen}, \&
  {Nordstrom}}]{mayor1996}
{Mayor}, M., {Duquennoy}, A., {Udry}, S., {Andersen}, J., \& {Nordstrom}, B.
  1996, in Astronomical Society of the Pacific Conference Series, Vol.~90, The
  Origins, Evolution, and Destinies of Binary Stars in Clusters, ed. E.~F.
  {Milone} \& J.-C. {Mermilliod}, 190

\bibitem[{{Mayor} {et~al.}(1997){Mayor}, {Meylan}, {Udry}, {Duquennoy},
  {Andersen}, {Nordstrom}, {Imbert}, {Maurice}, {Prevot}, {Ardeberg}, \&
  {Lindgren}}]{mayor1997}
{Mayor}, M., {Meylan}, G., {Udry}, S., {et~al.} 1997, \aj, 114, 1087

\bibitem[{{Monelli} {et~al.}(2003){Monelli}, {Pulone}, {Corsi}, {Castellani},
  {Bono}, {Walker}, {Brocato}, {Buonanno}, {Caputo}, {Castellani}, {Dall'Ora},
  {Marconi}, {Nonino}, {Ripepi}, \& {Smith}}]{monelli2003}
{Monelli}, M., {Pulone}, L., {Corsi}, C.~E., {et~al.} 2003, \aj, 126, 218

\bibitem[{{Norris} \& {Da Costa}(1995)}]{norris_dacosta1995}
{Norris}, J.~E. \& {Da Costa}, G.~S. 1995, \apj, 447, 680

\bibitem[{{Norris} {et~al.}(1997){Norris}, {Freeman}, {Mayor}, \&
  {Seitzer}}]{norris1997}
{Norris}, J.~E., {Freeman}, K.~C., {Mayor}, M., \& {Seitzer}, P. 1997, \apjl,
  487, L187

\bibitem[{{Norris} {et~al.}(1996){Norris}, {Freeman}, \&
  {Mighell}}]{norris1996}
{Norris}, J.~E., {Freeman}, K.~C., \& {Mighell}, K.~J. 1996, \apj, 462, 241

\bibitem[{{Pancino} {et~al.}(2000){Pancino}, {Ferraro}, {Bellazzini}, {Piotto},
  \& {Zoccali}}]{pancino2000}
{Pancino}, E., {Ferraro}, F.~R., {Bellazzini}, M., {Piotto}, G., \& {Zoccali},
  M. 2000, \apjl, 534, L83

\bibitem[{{Pancino} {et~al.}(2007){Pancino}, {Galfo}, {Ferraro}, \&
  {Bellazzini}}]{pancino2007}
{Pancino}, E., {Galfo}, A., {Ferraro}, F.~R., \& {Bellazzini}, M. 2007, \apjl,
  661, L155

\bibitem[{{Pancino} {et~al.}(2003){Pancino}, {Seleznev}, {Ferraro},
  {Bellazzini}, \& {Piotto}}]{pancino2003}
{Pancino}, E., {Seleznev}, A., {Ferraro}, F.~R., {Bellazzini}, M., \& {Piotto},
  G. 2003, \mnras, 345, 683

\bibitem[{{Piotto} {et~al.}(2005){Piotto}, {Villanova}, {Bedin}, {Gratton},
  {Cassisi}, {Momany}, {Recio-Blanco}, {Lucatello}, {Anderson}, {King},
  {Pietrinferni}, \& {Carraro}}]{piotto2005}
{Piotto}, G., {Villanova}, S., {Bedin}, L.~R., {et~al.} 2005, \apj, 621, 777

\bibitem[{{Rest} {et~al.}(2014){Rest}, {Scolnic}, {Foley}, {Huber}, {Chornock},
  {Narayan}, {Tonry}, {Berger}, {Soderberg}, {Stubbs}, {Riess}, {Kirshner},
  {Smartt}, {Schlafly}, {Rodney}, {Botticella}, {Brout}, {Challis}, {Czekala},
  {Drout}, {Hudson}, {Kotak}, {Leibler}, {Lunnan}, {Marion}, {McCrum},
  {Milisavljevic}, {Pastorello}, {Sanders}, {Smith}, {Stafford}, {Thilker},
  {Valenti}, {Wood-Vasey}, {Zheng}, {Burgett}, {Chambers}, {Denneau}, {Draper},
  {Flewelling}, {Hodapp}, {Kaiser}, {Kudritzki}, {Magnier}, {Metcalfe},
  {Price}, {Sweeney}, {Wainscoat}, \& {Waters}}]{rest14}
{Rest}, A., {Scolnic}, D., {Foley}, R.~J., {et~al.} 2014, \apj, 795, 44

\bibitem[{{Rest} {et~al.}(2005){Rest}, {Stubbs}, {Becker}, {Miknaitis},
  {Miceli}, {Covarrubias}, {Hawley}, {Smith}, {Suntzeff}, {Olsen}, {Prieto},
  {Hiriart}, {Welch}, {Cook}, {Nikolaev}, {Huber}, {Prochtor}, {Clocchiatti},
  {Minniti}, {Garg}, {Challis}, {Keller}, \& {Schmidt}}]{rest05}
{Rest}, A., {Stubbs}, C., {Becker}, A.~C., {et~al.} 2005, \apj, 634, 1103

\bibitem[{{Saha} {et~al.}(2010){Saha}, {Olszewski}, {Brondel}, {Olsen},
  {Knezek}, {Harris}, {Smith}, {Subramaniam}, {Claver}, {Rest}, {Seitzer},
  {Cook}, {Minniti}, \& {Suntzeff}}]{saha2010}
{Saha}, A., {Olszewski}, E.~W., {Brondel}, B., {et~al.} 2010, \aj, 140, 1719

\bibitem[{{Schechter} {et~al.}(1993){Schechter}, {Mateo}, \&
  {Saha}}]{schechter1993}
{Schechter}, P.~L., {Mateo}, M., \& {Saha}, A. 1993, \pasp, 105, 1342

\bibitem[{{Schlafly} \& {Finkbeiner}(2011)}]{schlafly2011}
{Schlafly}, E.~F. \& {Finkbeiner}, D.~P. 2011, \apj, 737, 103

\bibitem[{{Scolnic} {et~al.}(2015){Scolnic}, {Casertano}, {Riess}, {Rest},
  {Schlafly}, {Foley}, {Finkbeiner}, {Tang}, {Burgett}, {Chambers}, {Draper},
  {Flewelling}, {Hodapp}, {Huber}, {Kaiser}, {Kudritzki}, {Magnier},
  {Metcalfe}, \& {Stubbs}}]{scolnic2015}
{Scolnic}, D., {Casertano}, S., {Riess}, A., {et~al.} 2015, \apj, 815, 117

\bibitem[{{Sirianni} {et~al.}(2005){Sirianni}, {Jee}, {Ben{\'{\i}}tez},
  {Blakeslee}, {Martel}, {Meurer}, {Clampin}, {De Marchi}, {Ford}, {Gilliland},
  {Hartig}, {Illingworth}, {Mack}, \& {McCann}}]{sirianni2005}
{Sirianni}, M., {Jee}, M.~J., {Ben{\'{\i}}tez}, N., {et~al.} 2005, \pasp, 117,
  1049

\bibitem[{{Sollima} {et~al.}(2009){Sollima}, {Bellazzini}, {Smart}, {Correnti},
  {Pancino}, {Ferraro}, \& {Romano}}]{sollima2009}
{Sollima}, A., {Bellazzini}, M., {Smart}, R.~L., {et~al.} 2009, \mnras, 396,
  2183

\bibitem[{{Sollima} {et~al.}(2007{\natexlab{a}}){Sollima}, {Ferraro}, \&
  {Bellazzini}}]{sollima2007c}
{Sollima}, A., {Ferraro}, F.~R., \& {Bellazzini}, M. 2007{\natexlab{a}},
  \mnras, 381, 1575

\bibitem[{{Sollima} {et~al.}(2007{\natexlab{b}}){Sollima}, {Ferraro},
  {Bellazzini}, {Origlia}, {Straniero}, \& {Pancino}}]{sollima2007a}
{Sollima}, A., {Ferraro}, F.~R., {Bellazzini}, M., {et~al.} 2007{\natexlab{b}},
  \apj, 654, 915

\bibitem[{{Sollima} {et~al.}(2005){Sollima}, {Ferraro}, {Pancino}, \&
  {Bellazzini}}]{sollima2005a}
{Sollima}, A., {Ferraro}, F.~R., {Pancino}, E., \& {Bellazzini}, M. 2005,
  \mnras, 357, 265

\bibitem[{{Suntzeff} \& {Kraft}(1996)}]{suntzeff1996}
{Suntzeff}, N.~B. \& {Kraft}, R.~P. 1996, \aj, 111, 1913

\bibitem[{{van de Ven} {et~al.}(2006){van de Ven}, {van den Bosch}, {Verolme},
  \& {de Zeeuw}}]{vandeven2006}
{van de Ven}, G., {van den Bosch}, R.~C.~E., {Verolme}, E.~K., \& {de Zeeuw},
  P.~T. 2006, \aap, 445, 513

\bibitem[{{van Leeuwen} {et~al.}(2000){van Leeuwen}, {Le Poole}, {Reijns},
  {Freeman}, \& {de Zeeuw}}]{vanLeeuwen2000}
{van Leeuwen}, F., {Le Poole}, R.~S., {Reijns}, R.~A., {Freeman}, K.~C., \& {de
  Zeeuw}, P.~T. 2000, \aap, 360, 472

\bibitem[{{Villanova} {et~al.}(2007){Villanova}, {Piotto}, {King}, {Anderson},
  {Bedin}, {Gratton}, {Cassisi}, {Momany}, {Bellini}, {Cool}, {Recio-Blanco},
  \& {Renzini}}]{villanova2007}
{Villanova}, S., {Piotto}, G., {King}, I.~R., {et~al.} 2007, \apj, 663, 296

\end{thebibliography}

\end{document}